\documentclass[11pt,preprint,graphicx]{aastex}

\def\etal{\it et al. \rm }
\begin{document}

\title{The Age of Cluster Galaxies from Continuum Colors}

\author{Karl Rakos}
\affil{Institute for Astronomy, University of Vienna, A-1180, Wien, Austria;
karl.rakos@chello.at}

\author{James Schombert}
\affil{Department of Physics, University of Oregon, Eugene, OR 97403;
js@abyss.uoregon.edu}

\author{Andrew Odell}
\affil{Department of Physics and Astronomy, Northern Arizona University, Box 6010,
Flagstaff, AZ 86011; andy.odell@nau.edu}

\begin{abstract}

We determine the age of 1,104 early-type galaxies in eight rich clusters
($z = 0.0046$ to $0.175$) using a new continuum color technique.  We find
that galaxies in clusters divide into two populations, an old population
with a mean age similar to the age of the Universe (12 Gyrs) and a younger
population with a mean age of 9 Gyrs.  The older population follows the
expected relations for mass and metallicity that imply a classic monolithic
collapse origin.  Although total galaxy metallicity is correlated with
galaxy mass, it is uncorrelated with age.  It is impossible, with the
current data, to distinguish between a later epoch of star formation,
longer duration of star formation or late bursts of star formation to
explain the difference between the old and young populations.  However, the
global properties of this younger population are correlated with cluster
environmental factors, which implies secondary processes, post-formation
epoch, operate on the internal stellar population of a significant
fraction of cluster galaxies.  In addition, the mean age of the oldest
galaxies in a cluster are correlated with cluster velocity dispersion
implying that galaxy formation in massive clusters begins at earlier epochs
than less massive clusters.

\end{abstract}


\section{INTRODUCTION}

The age of the stars in a galaxy is a key parameter not only to
cosmological models (i.e. the epoch of initial star formation being a lower
limit to the age of the Universe), but also to our understanding of the
star formation history of galaxies.  The age of stellar populations in
galaxies guides our understanding of likely formation scenarios, e.g.
monolithic versus hierarchical models (Kauffmann, White \& Guiderdoni
1993).  However, even if a galaxy's stellar population age is decoupled
from the formation epoch, the history of star formation and chemical
evolution are determined primarily by the age of the stars and this value
still serves as one of the most important variables in understanding galaxy
evolution (Tinsley 1980).

Of course, since Nature is cruel (de Sade 1797), by far the hardest
parameter to extract from a galaxy's spectral energy distribution is the
age of the underlying stellar population.  The reason is mostly technical,
in that there is no single color or spectral index that uniquely measures
age (as derived from the turnoff point for the stellar population).  Thus,
one is required to deduce a mean age based on a series of observables, such
as the combined $<$Fe$>$ versus $H\beta$ index (Worthey 1994) or $UV$
versus optical colors (Kaviraj \etal 2006), guided by synthetic models
based on stellar physics.  These models have only recently achieved a
necessary level of success to transform from isochrone fits to synthetic
colors, i.e. all the relevant stellar physics is correctly incorporated (Yi
\etal 2003, Schiavon 2007).

Techniques to measure metallicity and age have divided into two paths, use
of line indices (e.g. the Lick/IDS system, Gonzalez 1993, Trager \etal
2000) versus continuum colors (Rakos \& Schombert 2005).  These two
techniques are, effectively, attempting to determine age and metallicity
through slightly different astrophysical phenomenon.  Line indices can
trace the metallicity of a system by directly measuring the abundance of
various element lines (although still influenced by the age of the system,
Trager \etal 2000).  Early work (Burstein \etal 1984) focused on easily
measurable lines such as Mg$_2$ and Ca $H$ and $K$ (Faber 1973).  These
individual lines contain a wealth of information on the chemical history of
stellar populations and the process of nucleosynthesis (Pagel 2001).
However, extracting the mean metallicity of a stellar population, as
expressed by [Fe/H], can be problematic if the ratio of these elements to
Fe varies.  For example, Type Ia SN produce more Fe to $\alpha$ elements
than Type II, but require over 1 Gyr to evolve to the detonation stage such
that the $\alpha$/Fe ratio decreases with time (Maraston \etal 2003).
Thus, models which do not include these ratio variations risk incorrect
estimates for color and line indice changes.  Improved technology has meant
that, for a large number of galaxies, the relevant metallicity value, e.g.
[Fe/H], can be determined directly from Fe lines themselves (e.g. Fe4383
and Fe5015, Gonzalez 1993, Trager \etal 2000, S\'{a}nchez-Bl\'{a}zquez
\etal 2006).  As Fe is the primary source of electrons in stellar
atmospheres, it is a trivial conversion from [Fe/H] to total metallicity,
$Z$ or [M/H].

Age is determined in line indice studies, primarily, through the $H\beta$
index (see Schiavon 2007 for a detailed discussion).  While this age
determination is still model dependent, the separation in parameter space
is clear (e.g. MgFe versus $H\beta$).  Notably, in contrast to our
expectation of old stellar populations from interpretation of the
color-magnitude relation (Faber 1973), line indices studies find a high
fraction of early-type galaxies with young ages (less than 7 Gyrs) in
agreement with certain CDM frameworks for galaxy formation (Kauffmann,
White \& Guiderdoni 1993) that predict extended star formation histories
(Kuntschner \& Davies 1998, Trager \etal 2000).  Later re-analysis, with
refined spectral energy distribution (SED) models, finds that only 25\% of
the galaxies in previous samples have ages less than 7 Gyrs (Schiavon
2007).  However, even this fraction is unexpectedly high given the
tightness of the color-magnitude relation and the expected increased
scatter due to the influence of age on color (Bower \etal 1992).

Narrow band continuum colors, used in this work, approach the age and
metallicity determination problem from a different direction.  They measure
the effects of metallicity, in composite stellar populations, by the change
in color of the red giant branch (RGB) with [Fe/H].  Age effects are
measured by color changes produced in the shifting of the turnoff point
(Tinsley 1980).  This type of data collection (i.e. imaging) has the
advantage of being faster (therefore, requiring less observing time or
capable of penetrating to fainter luminosities than spectroscopy) and
provides spatial information.  However, as with line indice work, this
technique is also dependent on the use of SED models to interpret the
resulting colors (see Steindling, Brosch \& Rakos 2001 for an in-depth
analysis of the technique).  We have recently adopted a more reliable
principal component (PC) technique that is tied to the galactic globular
age and metallicity scale (Rakos \& Schombert 2005) which has resulted in a
higher degree of accuracy in age determination than our previous papers.

The goal of this paper is three-fold.  First, we briefly review our
technique, its strengths and weaknesses, our multi-metallicity models and,
most importantly, where the key uncertainties lie.  Second, we will present
our most recent compendium of ages and metallicities for cluster galaxies
based on recent observations of four new clusters combined with data from
four clusters previously published in the literature, now re-analyzed with
our new techniques.  Third, using the age and [Fe/H] values extracted from
our colors, we will explore the trends and correlations as constrained by
the limitations of our technique.  Throughout this paper we use the
Benchmark cosmological parameters of $\Omega_M=0.3$, $\Omega_{\Lambda}=0.7$
and $H_o=75$.

\section{OBSERVATIONS}

\subsection{Cluster Photometry}

A total of eight clusters were used for this study.  In order of increasing
redshift, they are Fornax ($z=0.0046$), Coma ($z=0.0231$), A119
($z=0.0442$), A400 ($z=0.0244$), A539 ($z=0.0284$), A779 ($z=0.0225$),
A1185 ($z=0.0325$) and A2218 ($z=0.175$).  Four clusters (Fornax, Coma,
A1185 and A2218) were published in our earlier studies (Rakos, Schombert \&
Odell 2007, Rakos \& Schombert 2005, Odell, Schombert \& Rakos 2002, Rakos
\etal 2001).  The remaining four (A119, A400, A539, A779) were obtained in
the last few observing seasons using the 2.3m Bok telescope of the Steward
Observatory located at Kitt Peak Arizona.  The image device was 90prime, a
prime focus wide-field imager using a mosaic of four 4k by 4k CCDs
resulting in a field of view of 1.16 by 1.16 degrees and a plate scale of
0.45"/pixel (Williams \etal 2004).

\begin{figure}
\centering
\includegraphics[scale=0.95]{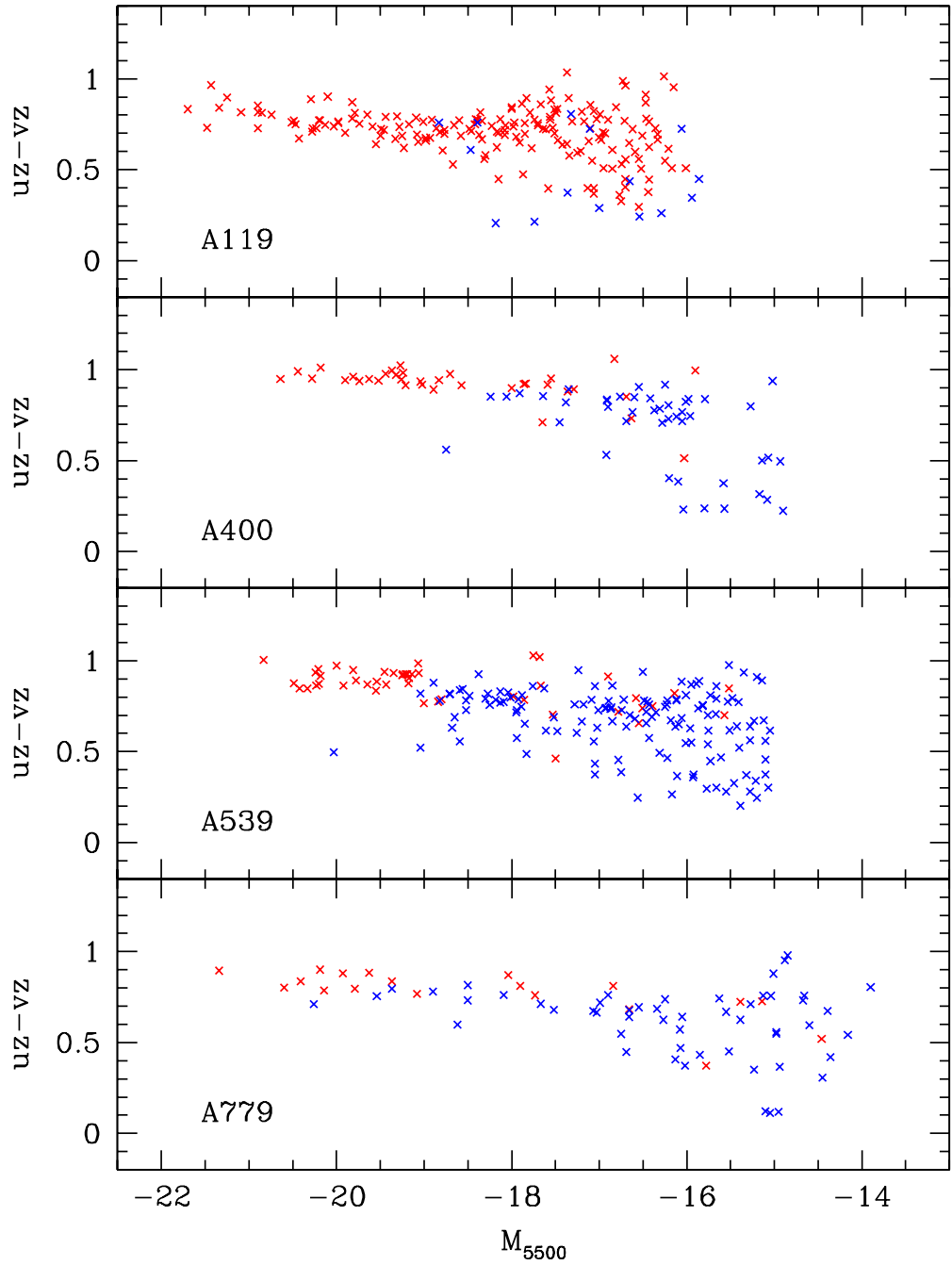}
\caption{The color-magnitude diagram of our four new clusters in our bluest
filter , $uz-vz$.  Ages, deduced by our PCA method, are indicated as
greater than 10 Gyrs (marked in red) versus those less than 10 Gyrs (marked
in blue).
}
\end{figure}

The filter system for this study is a modified Str\"omgren ($uvby$) system,
the same as used in our distant cluster studies (Rakos \& Schombert
1995).  The modified Str\"omgren system (herein called $uz,vz,bz,yz$) is
altered in the sense that the filters are slightly narrower (by 20\AA) than
the original Str\"omgren system and the $uz$ filter is shifted 30\AA\ to the
red of its original central wavelength.  The system we use herein is called
the $uz,vz,bz,yz$ system to differentiate it from the original $uvby$
system due to our small changes plus our filters are specific to the rest
frame of the cluster that is being studied  (this avoids k-corrections and
allows for photometric membership to eliminate foreground and background
galaxies).  The $uz,vz,bz,yz$ system covers three regions in the near-UV
and blue portion of the spectrum that make it a powerful tool for the
investigation of stellar populations in SSP's (simple stellar population),
such as star clusters, or composite systems, such as galaxies.  The first
region is longward of 4600\AA, where the influence of absorption lines
is small.  This is characteristic of the $bz$ and $yz$ filters
($\lambda_{eff}$ = 4675\AA\ and 5500\AA), which produce a temperature color
index, $bz-yz$.  The second region is a band shortward of 4600\AA, but
above the Balmer discontinuity. This region is strongly influenced by metal
absorption lines (i.e. Fe, CN) particularly for spectral classes F to M,
which dominate the contribution of light in old stellar populations.  This
region is exploited by the $vz$ filter ($\lambda_{eff} = 4100$\AA).  The
third region is a band shortward of the Balmer discontinuity or below the
effective limit of crowding of the Balmer absorption lines.  This region is
explored by the $uz$ filter ($\lambda_{eff} = 3500$\AA).  All the filters
are sufficiently narrow (FWHM = 200\AA) to sample regions of the spectrum
unique to the various physical processes of star formation and metallicity
(see Rakos \etal 2001 for a fuller description of the color system and its
behavior for varying populations).

The reduction procedures have been published in Rakos, Maindl \& Schombert
(1996) and references therein. The photometric system is based on the
theoretical transmission curves of filters (which can be obtained from the
authors) and the spectra of spectrophotometric standard stars published in
the literature (Massey \& Gronwall, 1990).  The convolution of the
transmission curves and the spectra of the standard stars produces
theoretical flux values for color indices of the standard stars corrected
for all light losses in the equipment and the specific sensitivity of the
CCD camera.  Magnitudes were measured on the co-added images using standard
IRAF procedures and are based for brighter objects on metric apertures set
at 32 kpc for cosmological parameters of $H_o=75$ km s$^{-1}$ Mpc$^{-1}$
and the Benchmark cosmology ($\Omega_m=0.3$, $\Omega_\Lambda=0.7$).  For
fainter objects, the apertures were adapted to deliver the best possible
signal to noise ratio, but always using the same aperture for all four
filters.  Color indices are formed from the magnitudes: $uz-vz$, $bz-yz$,
$vz-yz$, and $mz$ [$=(vz-bz)-(bz-yz)$].  Galactic extinction was
determined, and corrected for, in the standard fashion.  Typical errors
were 0.03 mag in $vz-yz$ at the bright end of the sample and 0.08 mag at
the faint end.  The color-magnitude diagrams for the four newest cluster
are shown in Figure 1 for the $uz-vz$ filters.  The characteristics and
slopes of these diagrams are identical to previous color-magnitude
diagrams.

\subsection{Measuring Age in Galaxies}

Determination of age by line indices focuses on the particular lines
emitted by stars near the turnoff point, in this case Balmer lines from A
(and hotter) stars.  The strongest Balmer line, which is relatively free of
contaminating emission from interstellar gas, is $H\beta$.  The use of
$H\beta$ is well documented (Trager \etal 2000, Kuntschner \etal 2001,
Poggianti \etal 2001, Gallazzi \etal 2005, Thomas \etal 2005,
S\'{a}nchez-Bl\'{a}zquez \etal 2006, Schiavon 2007) and has the additional
advantage that comparison to higher order Balmer lines allows for
determination of not only mean age, but duration of star formation
(Schiavon 2007, see also Prochaska \etal 2007 for problems with this
technique).

Continuum colors determine age by measuring the turnoff point, in this
case, the mean color of those stars in the near-blue region of the
spectrum.  Clearly, there is competition with the changes in color due to
metallicity effects (the so-called age-metallicity degeneracy problem,
Worthey 1999), but, this conflict can be resolved by either a longer
baseline in color (e.g. near-UV for age, near-IR for metallicity as the
turnoff stars have a stronger influence on blue colors than RGB stars) or
by judicious choice of a narrow band continuum colors, as is the case in
our work (Rakos \etal 2001).

While there are technical challenges for both line indice and continuum
color measurements, both techniques have a primary difficulty in
determining galaxy age in that, for a composite stellar population (one
composed of stars with a range of ages), the age value measured is, in
fact, a luminosity weighted mean age.  Even if the underlying stars have a
normal distribution in age, such that the mean age reflects the average,
there still exists extremely relevant information contained in the spread
of ages (i.e. the duration of initial star formation or later phases of
star formation).  And, galaxies must be composed of more than a simple
stellar population (SSP, i.e. singular age and metallicity such as a
globular cluster) for various reasons.  Observationally, the existence of
color gradients in ellipticals (see review by Kormendy \& Djorgovski 1989)
is perhaps the earliest evidence to this fact as all early-type galaxies
have stellar population differences from their cores to halos.  All
indications are that this change in color is due to metallicity effects
(S\'{a}nchez-Bl\'{a}zquez \etal 2007), but regardless of whether age or
metallicity dominates, it is clear that galaxies are composed of stellar
populations with a range of these values.  HST imaging has resolved the
stellar populations in several nearby ellipticals (Grebel 2003, Tosi 2007),
all demonstrating the existence of a distribution of metallicities, but the
question of a distribution of ages remains unresolved (Grebel 2005).  Our
own most recent work on a comparison of [Fe/H] estimates between line
indice work and our continuum colors indicates that the distribution of
metallicity in ellipticals must be broad and is similar to pre-enrichment
infall models (Schombert \& Rakos 2008).

Since galaxies are composed of multiple stellar populations, then the
observed age, being an integrated value, can be arrived at by various paths
of star formation.  For example, for a calculated age that is a luminosity
weighted mean age of 8 Gyrs can be obtained by a late
epoch of star formation (5 Gyrs later than the oldest galaxies) or by an
extended duration of star formation (i.e. going from 12 Gyrs ago up to 4
Gyrs from the current epoch to an average of 8 Gyrs).  In addition to
varying durations of star formation, an intermediate aged population might
exist having its origin from a recent, weak burst of star formation (Serra
\& Trager 2007, Trager \etal 2005).  Thus, a secondary clock, such as
the ratio of $\alpha$ elements to Fe (a measure of the contribution by Type
Ia SN to the chemical enrichment on timescales of Gyrs) will be required to
constrain various age distribution scenarios (see \S3.4).

\subsection{Continuum Color Method of Age Determination}

Our technique to determine age and metallicity using narrow band continuum
colors is outlined, in great detail, in Rakos \& Schombert (2005).  Our
method differs from line studies by measuring the effects of varying age
and metallicity on the shape of selected parts of a galaxy spectrum.  The
underlying phenomenon that we capture are 1) varying metallicity reflects
into the position of the RGB (due to the stronger effect of line blanketing
in evolved stars) and 2) age is reflected in the position of the turnoff
point.  The turnoff point changes in color slowly with age after 3 Gyrs
(Schulz \etal 2002) and, for this reason, our technique fails for younger
populations, although we do not expect this to be a problem for red 
early-type galaxies in rich clusters.  In any case, star forming galaxies
signal their young nature through several of our filters and, while we
can not determine their exact ages, we can exclude them from our samples.

Our use of narrow band continuum colors (a modified $uvby$ Str\"omgren
system) is motivated by three factors; 1) narrow filters select smaller
regions of a galaxy's spectrum, regions which show different sensitivities
to changes in the RGB and turnoff point and, thus, facilitate the breaking
of the age-metallicity degeneracy, 2) there exist improved population
models (Schulz \etal 2002) which sample the region of age-metallicity
parameter space useful for the analysis of old, red galaxies and 3) these
models are now tied, accurately, to the globular cluster ages and
metallicities resolving a long standing dilemma for stellar population
studies.  The Schulz \etal models are a significant improvement over past
work with respect to our near-UV and blue colors as they 1) include the
recent isochrones from the Padova group which have a stronger blue HB
contribution (important for even a small old, metal-poor contribution to
the galaxy light), 2) use more recent stellar atmosphere spectra which,
again, are refined to better sample the blue region of a galaxy's spectrum
and 3) are in better agreement with globular cluster $uvby$ colors than
previous work.  Our use of the $uz,vz,bz,yz$ filters to determine age and
metallicity culminated in Rakos \& Schombert (2005) where we develop a
principal component methodology for determining these two parameters in
composite systems, such as galaxies.

Briefly, our principal component technique derives metallicity and age
values using the observed narrow band colors and the knowledge of the
behavior of the PC surface as mapped by the Schulz \etal SED models.  A
grid search method is used in an iterative fashion to locate the pair of
age and metallicity that minimize the distance between calculated PC values
with model PC values.  This technique has the advantage of using the
information from all four filters simultaneously while allowing flexible
use of various SED models and yet is still tied to the globular cluster
age/metallicity system.

This method of determining galaxy age based on continuum narrow band colors
has several caveats.  First, this measurement is the mean age of the
underlying stellar population, not the time from galaxy formation.  
We will argue in a later section that, for the oldest galaxies in the
sample, this stellar population age must coincide with the formation epoch.
However, for galaxies with stellar population ages several Gyrs less than
the age of the Universe, there is a broad window of time for the actual
assembly time of the galaxy (stars plus dark matter).

Another limitations to our technique arises from the models themselves.
For example, the models are restricted to [Fe/H] between $-$1.7 and $+$0.5
plus ages between 3 and 15 Gyrs and, thus, our extracted values are
constrained by mild extrapolation of these limits.  With respect to ages
for cluster galaxies, this limitation will not impact our results as we can
isolate galaxies with ongoing star formation, or have large fractions of
young stars, by their extreme colors and/or strong emission lines.  These
galaxies are identified as their colors lie outside the color parameter
space of the SED models and, thus, are not submitted for age/metallicity
analysis.  These objects consisted of less than 4\% of our cluster sample.
The remaining galaxies have colors which indicate they have been quiescent
on timescales of Gyrs, and are ideal candidates for study by our technique.
In addition, with respect to metallicity, there is a strong expectation,
based on previous work, that cluster galaxies contain a majority of their
stars with mean [Fe/H] values greater than globular clusters, removing any
problems of fitting the low metallicity end of the models.

Our resulting ages and metallicities are thus based on the PC analysis of
three narrow band colors ($uz-vz$, $vz-yz$, $bz-yz$, Rakos \& Schombert
1995).  Our ages are deduced based on SSP values (i.e. we assume no spread
in age within a galaxy's stellar population) but we test that assumption in
\S3.4.  We do adjust our metallicity values based our work comparing line
indice [Fe/H] values to colors produced by a multi-metallicity model.  This
adjustment is linear and based on our `push' model of chemical enrichment
(discussed in Schombert \& Rakos 2008).  Briefly, the multi-metallicity
models attempt to reproduce the galaxy colors using a real distribution of
internal metallicities (i.e. a chemical evolution model) rather than
singular SSP.  This is done by taking the SSP's for a range of
metallicities (follow some model of the distribution of [Fe/H]) and summing
the colors weighted by the luminosity per mass of the SSP's.  Since lower
metallicity SSP's contribute more blue light than higher metallicities,
then the luminosity weighted value for [Fe/H] will tend to be lower than
the [Fe/H] from a numerical average of the SSP's.  The 'push' model is a
simple closed box [Fe/H] distribution pushed at low [Fe/H] values to
simulate the G-dwarf problem (i.e. initial enrichment).  Due to the varying
luminosity contributions by each bin of metallicity, the calculated value
for [Fe/H] must be adjusted to produced the mass averaged value
of metallicity, $<$Fe/H$>$.  While this adjustment is negligible at low
metallicities ([Fe/H] $<$ $-$1.5), there are increases in [Fe/H] by 0.1 dex
at values greater than solar, and it is this metallicity that we use in our
following discussions.

\section{DISCUSSION}

\subsection{Galaxy Age from Narrow Band Colors}

A graphical summary of our deduced mean ages and [Fe/H] for 1,104 galaxies
in eight clusters (in order of increasing redshift: Fornax, A779, Coma,
A400, A539, A1185, A119, A2218) is shown in Figures 2, 3 and 4, plots of
log $\tau$ (Gyrs), [Fe/H] and stellar mass (log $M_*/M_{\sun}$).  Distance
is derived using the Benchmark cosmology and stellar mass is calculated
from the $M_{5500}$ luminosity assuming a M/L of 3 (Bender, Burstein \&
Faber 1992).  Since our sample is composed primarily of early-type
galaxies, the amount of gas is minimal and, therefore, the stellar mass
reflects the total baryonic mass of the galaxy.  Typical uncertainties,
defined mostly by the SED models, are 0.5 Gyrs in age, 0.2 dex in [Fe/H]
(although this will increase at lower luminosities due to a combination of
higher errors in the observations and running against the limits of the SED
models).  The uncertainties are represented by the greyscale image in each
plot where each data point is taken to be a 2D gaussian of width given by
the errors above.  Each datapoint's contribution per bin is summed to
produce the greyscale intensity, normalized to the total sample size.

\begin{figure}
\centering
\includegraphics[scale=0.95]{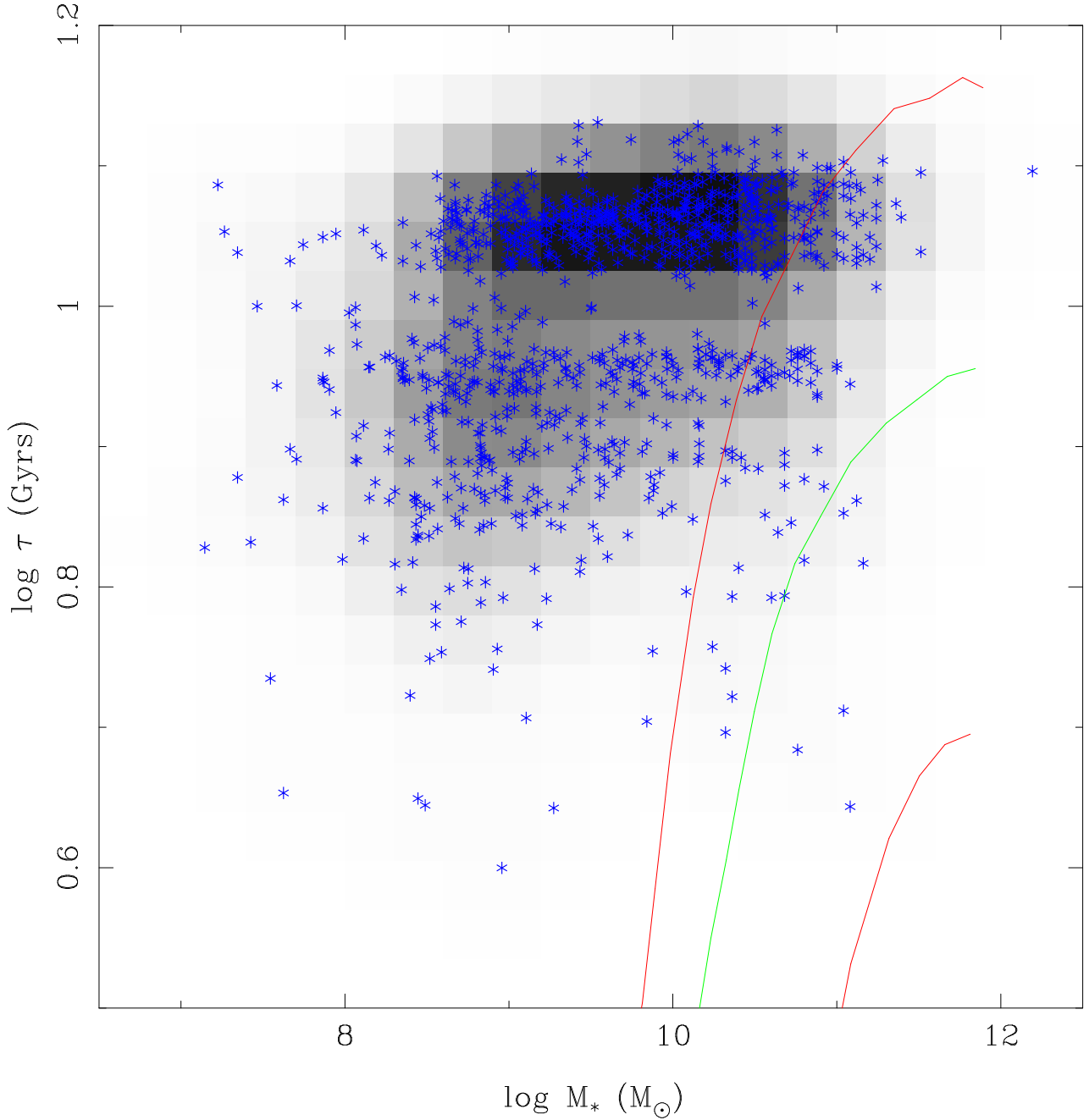}
\caption{Mean galaxy age versus galaxy stellar mass (in solar units) for
all 1,104 galaxies in our cluster sample.  Cluster galaxies divide into two
distinct populations, an old population with mean ages of 11.5 Gyrs and a
younger population (the gap population) with a mean age of 9 Gyrs.  There
is a slight tendency for the oldest galaxies to decrease in age with lower
mass.  The statistically significant decrease in age with stellar mass for
the gap population is driven by the increasing number of young galaxies
($\tau <$ 7 Gyrs) for masses less than $10^9 M_{\sun}$).  The green line
displays the mean relation of age versus stellar mass from Gallazzi \etal
(2005), the red line displays the 1$\sigma$ range in the SDSS sample of age per mass
bin.  As is typical for line indices studies, the Gallazzi \etal study
finds sharply younger galaxy ages for decreasing mass, although we note that the SDSS sample
includes all galaxy types as compared to our strict cluster sample.
}
\end{figure}

All the key features noted over the decades relating to galaxy evolution
are visible.  For example, the mass-metallicity relation is evident in
Figure 3, although not the strictly linear relation as indicated by the
color-magnitude diagram for galaxies (Bower \etal 1992, Vazdekis \etal
2001).  There is some suggestion in this diagram of two populations, a high
mass, solar metallicity population with a strong correlation between mass
and $<$Fe/H$>$ plus a lower mass population with lower metallicities and a
much higher scatter.  Age does not appear to be a factor in the trend found
in Figure 3 as there are numerous old galaxies with [Fe/H] values around
$-$1 and plenty of younger galaxies with [Fe/H] near solar (although see
Figure 13 for the effect of age on the scatter in the color-magnitude
relation).  The only consistent trend with age is that all galaxies with
$<$Fe/H$>$ values less than $-$1.5 have ages less than 10 Gyrs.  It would
appear, since there is no evidence in Figures 3 or 4 that young galaxy age
leads to high metallicity values, that the epoch of star formation, or its
duration, is not a significant factor in determining the mean metallicity
of a galaxy.  Mass is the primary driver of metallicity as predicted by
infall models of chemical evolution (Pipino \& Matteucci 2004), although
probabily not the sole variable (see \S3.7).

Determination of the age of the stellar populations in galaxies is the
primary goal of this paper and a comparison in the range of galaxy ages we
find versus the range in ages by two recent line indice studies
(S\'{a}nchez-Bl\'{a}zquez \etal 2006 and Thomas \etal 2005) is found in
Figure 5.  This comparison is key to our discussion as the
S\'{a}nchez-Bl\'{a}zquez \etal and Thomas \etal studies use the Lick/IDS
system of age determination (i.e. the model grid interpretation of the
H$\beta$ versus $<$Fe$>$ or MgFe index).  As can be seen from Figure 5, the
distribution of galaxy ages differs dramatically between the Lick/IDS
technique and our method.  Both the S\'{a}nchez-Bl\'{a}zquez \etal and
Thomas \etal samples (only the high density sample is shown) find
significant numbers of galaxies with ages less than 7 Gyrs.  

Our technique is unable to calculate ages less than 3 Gyrs, however, those
objects still signal their young ages with star forming colors and are rare
in our cluster samples (Rakos, Maindl \& Schombert 1996).  We also note
that our technique finds that a majority of cluster galaxies have ages
greater than 10 Gyrs, but old galaxies are in the minority in the samples
that determine age by the Lick/IDS system.  For example, in Coma, we have
three galaxies in common with the Thomas \etal sample, NGC 4839, 4874 and
4889.  The Thomas \etal ages are 15.8, 2.4 and 2.8 Gyrs respectfully.  Our
continuum color ages are 11.6, 12.5 and 10.9 Gyrs for the same galaxies.
Our ages are more in agreement with the expectations of stellar population
evolution from intermediate redshift clusters (Rakos \& Schombert 1995),
where the Thomas \etal values would predict large numbers of bright cluster
ellipticals with star-forming colors, which is not seen.

\begin{figure}
\centering
\includegraphics[scale=0.95]{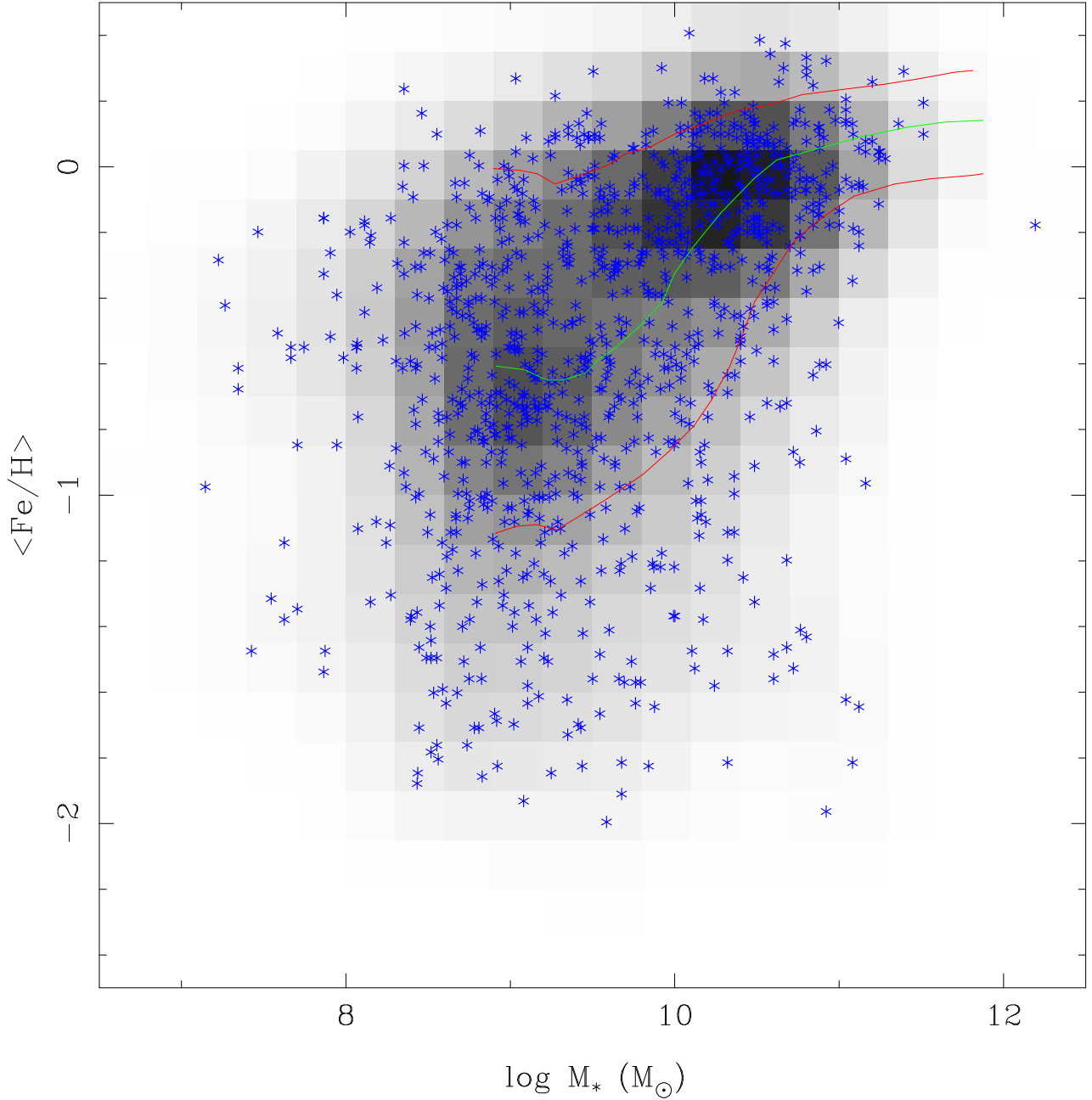}
\caption{Average metallicity, $<$Fe/H$>$, versus galaxy stellar mass.  The
average metallicity is the luminosity weighted metallicity from our PC
technique adjusted for the expected underlying average metallicity
(Schombert \& Rakos 2008).  The mass-metallicity relation is evident,
although, as noted by many previous studies, the scatter is higher than
observational uncertainties.  There are statistically significant numbers
of low mass galaxies with high metallicities and high mass galaxies with
intrinsically low [Fe/H] values.  Thus, while galaxy mass (depth of its
potential well) is the primary parameter in a galaxy's chemical evolution,
it is not its sole determinate.  The green line displays the mean relation
of [Fe/H] versus stellar mass from Gallazzi \etal (2005), the red line
displays the 1$\sigma$ range in the SDSS sample of metallicity per mass bin.  The
SDSS metallicity relation is in excellent agreement with our estimates.
}
\end{figure}

The differences between the Lick/IDS system method and other techniques of
age determination has been noted before (e.g. Bregman, Temi \& Bergman
2006).  In order to confirm the legitimacy of our age values, we have
turned to the third method mentioned in the Introduction, the use of longer
baseline broadband filters to exploit the sensitivity of near-IR colors to
the position of the RGB.  In this case, we can combine our narrow band
colors in the Coma cluster (Odell, Rakos \& Schombert 2002) with recent $K$
band data (Eisenhardt \etal 2007).  There are 16 galaxies in common with
their sample and ours, for which our near-blue color $vz-yz$ is plotted
against their near-IR color $V-K$ in Figure 6.  We label each galaxy data
point by its estimated age, using our technique, rounded to the nearest
Gyr.  In this Figure it is clear that galaxies with the oldest ages ($\tau
> 10$ Gyrs) separate from the younger systems ($7 < \tau < 10$ Gyrs).
Thus, the Coma galaxies with near-IR colors separate nicely in this
optical/near-IR color plane.  There is some overlap in ages, but within our
expected errors of 0.5 Gyrs.

This Figure also allows an opportunity to check various SED models tracks
to our age estimates and the absolute value of our age estimates (as our
technique is tied to the galactic globular cluster ages).  In this case, we
have chosen the most recent SED models from Bruzual \& Charlot (2003,
hereafter BC03).  Figure 6 displays the model color tracks over a range of
metallicities ([Fe/H] from $-$0.8 at $V-K=2.5$ to $+$0.3 at $V-K=3.5$ for
all the models) for a population of 8 and 12 Gyrs.  The dashed lines are
the raw SSP models from BC03, which do not match the colors of the Coma
sample.  This is not unexpected as galaxies are not composed of a single
metallicity population, as represented by the SSP's.  As discussed in
Schombert \& Rakos (2008), the color-magnitude relation and the line indice
versus color relations are best matched by a multi-metallicity model that
incorporates a modified infall model of chemical evolution.  These models,
referred to as the `push' models in Schombert \& Rakos (2008), are shown as
solid lines in Figure 6 and are an excellent match to the relative age
differences.  Thus, we conclude that our absolute and relative age
estimates conform to expectations from SED models.

\subsection{Cluster Ellipticals Ages}

There are several features to the age and metallicity distributions to our
sample that are common to all the clusters and visible in Figures 2, 3 and
4.  The first is that there clearly exist galaxies with old ages ($\tau
\approx$ 12 Gyrs) in all clusters.  This mean value is comparable to the
age of the Universe (i.e. these are the oldest galaxies) and this
population makes up the majority of cluster galaxies.

\begin{figure}
\centering
\includegraphics[scale=0.95]{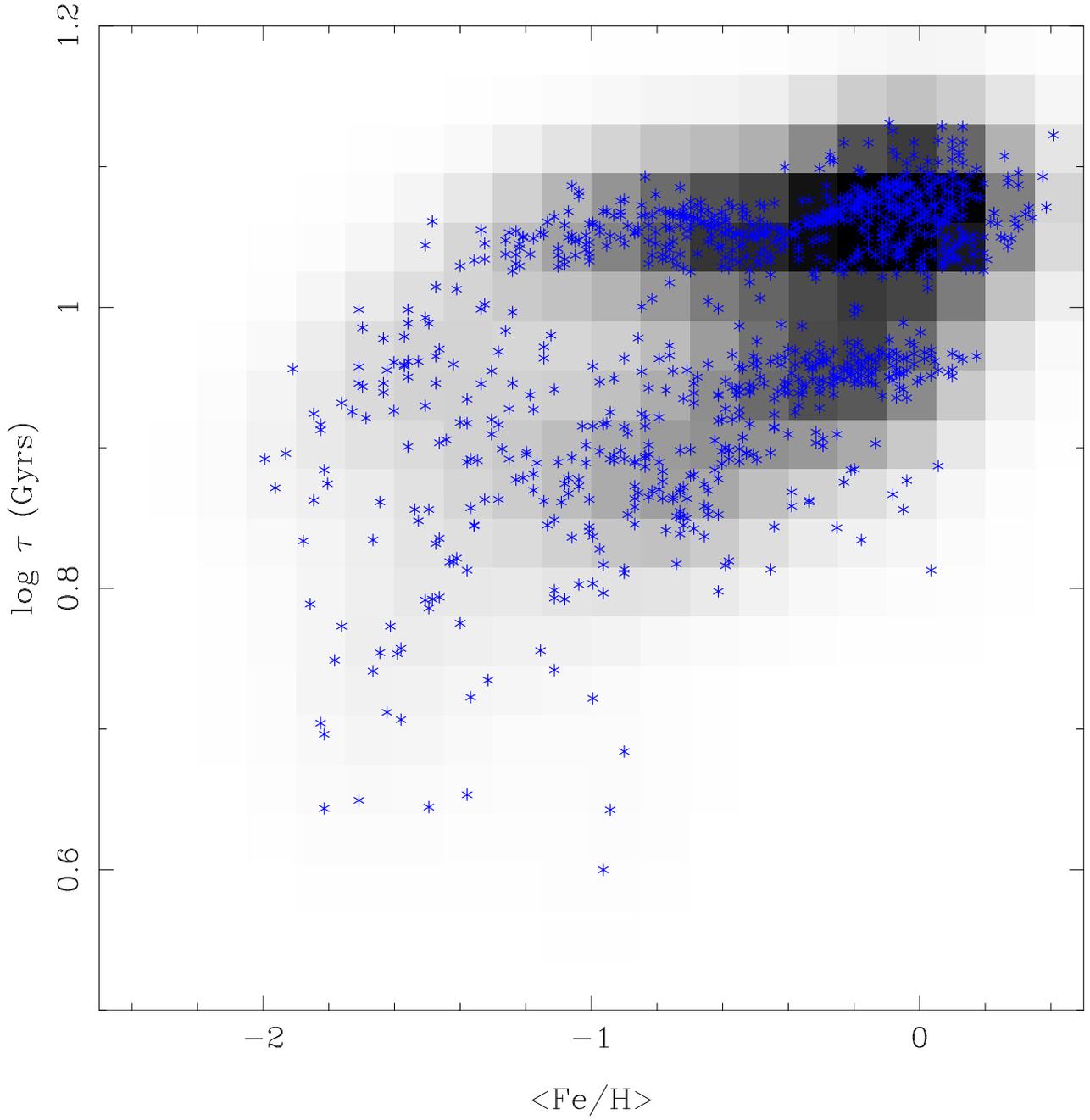}
\caption{Mean galaxy age versus mean metallicity.  The two populations are
still evident in this figure.  The oldest galaxies display a range of mean
metallicities (from $-$1.5 to $+$0.3).  The gap population displays a
correlation of decreasing metallicity with age.
}
\end{figure}

Second, the red cluster galaxies divided neatly into two
populations based on age, with a distinct gap in age between the
populations.  The older population has a mean of 11.5 Gyrs, the younger
population a mean age of 9.5 Gyrs.  The gap is wider than our estimated
error of 0.5 Gyrs, therefore, appears to be a real feature.  This gap is
not seen in studies that use the Lick/IDS, such as Thomas \etal (2005) or
S\'{a}nchez-Bl\'{a}zquez \etal (2006), but this may be due to small number
statistics.  The transition in age with mass is sharp in the Gallazzi
\etal (2005) work (see their Figure 9, also drawn in our Figure 2), which
involves over 175,000 galaxies from the SDSS database, but does not take on
a bimodality distribution as suggested by our data sample.  We note that
the SDSS sample includes all galaxy types, both field and cluster, whereas
our sample is a strict, high density, cluster population.  The inclusion of
star-forming disk systems would clearly deviate the correlation of age and
mass to younger ages.

\begin{figure}
\centering
\includegraphics[scale=0.95]{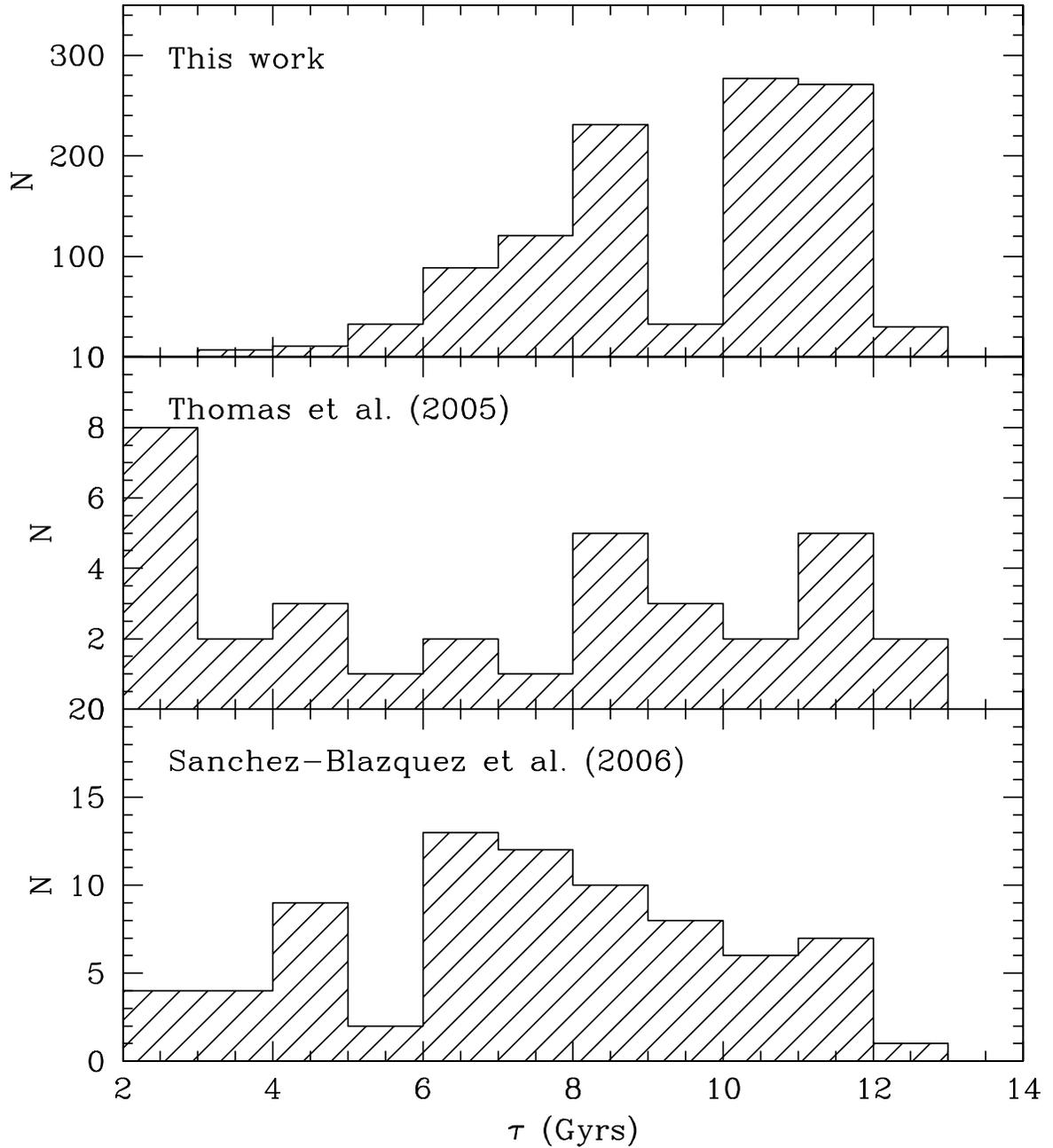}
\caption{A comparison between galaxy ages derived by continuum colors (this
work) and those derived from two Lick/IDS studies (Thomas \etal 2005, high
density sample, and S\'{a}nchez-Bl\'{a}zquez \etal 2006).  Typical for
spectroscopic studies, the Lick/IDS work finds a higher fraction of young
galaxies ($\tau <$ 7 Gyrs) compared to studies that use colors (Bregman,
Temi \& Bergman 2006).
}
\end{figure}

Lastly, there are a couple trends (i.e. weak correlations) with age.  In
general, brighter luminosity (higher mass) systems tend to be older.  Older
galaxies also tend to be more metal-rich (as measured by [Fe/H]).  However,
neither of these trends is strong suggesting that the path that a galaxy
evolves through to reach these global values of age and metallicity is
influenced by several factors not directly related to its total mass.  In addition,
the lack of sharp correlations indicates that the global values we measure,
such as luminosity-weighted age, may be composed of a more convoluted
underlying stellar population.  We will examine each of these features, and
combine our results here with previous results, in an attempt to untangle
our observations in the following sections.

\subsection{The Oldest Galaxies}

The oldest galaxies in our sample have mean ages between 11 and 12 Gyrs,
which is consistent with the assumption of a primordial epoch of initial
star formation and the age of the Universe under the Benchmark model (12.6
Gyrs).  While there exist galaxies with calculated ages of 13 Gyrs (only
1\% of the sample galaxies have ages greater than 13 Gyrs, none older than
13.5 Gyrs), they are consistent with the age of the Universe within the
errors of the age determination method.

Extremely old stellar populations in galaxies are not unexpected under
certain galaxy formation scenarios (e.g. monolithic collapse, Eggen,
Lynden-Bell \& Sandage 1962, Larson 1974, Arimoto \& Yoshii 1987), it is
somewhat surprising to find large numbers of purely old galaxies in our
sample as star formation has been observed in ellipticals between redshifts
of zero and 1 (Stanford \etal 2004).  Later star formation would push our
integrated age estimates to younger values (see \S3.4).

Two possible scenarios can resolve this dilemma.  The first is that the
observed high redshift star formation events may involve a very small
fraction of the total stellar population mass and, thus, their anomalous
colors have since faded below detection, overwhelmed by the colors of an
older stellar population.  The second is that the number of galaxies
undergoing these recent bursts have been overestimated.  There certainly
exist galaxies in our cluster samples with younger mean ages (see gap
population discussion below) and they may be the ancestors of these higher
redshift objects.

The existence of galaxies with ages nearly equal to the age of the Universe
places a series of important constraints on scenarios for galaxy formation.
While there are clearly evolutionary processes in play within rich clusters
(i.e., mergers and tidal collisions which induce later star formation,
Moore \etal 1996), it appears, from the properties of our reddest galaxies,
that there must exist a sizable fraction of cluster ellipticals whose
dominant stellar population dates back to the epoch of cluster formation
itself.  These are valuable inputs to our galaxy formation models as these
ages allow for no room in the time budget for later epochs of star
formation (which would lower the deduced mean age by our method) nor for a
very long duration of initial star formation after the time of galaxy
formation (which would also decrease the deduced mean age).  We note that
that Thomas, Maraston \& Bender (2002) find that the oldest galaxies in
their sample have the highest $\alpha$/Fe ratios, implying very short
initial durations of star formation in agreement with their old ages.
Short durations of star formation imply that the resulting chemical
evolution is direct, in agreement with our determination that the internal
metallicity distribution in ellipticals are bested described by simple
infall models (Schombert \& Rakos 2008).

At the very least, our oldest galaxies are a challenge to hierarchical
models where galaxies are assumed to assemble gradually with time by
mergers (Baugh, Cole \& Frenk 1996).  All the stars in these merging
systems would be required to have the same ages and chemical history to
explain the mass-metallicity relation (although perhaps the scatter in the
mass-metallicity diagram is a signature of past stochastic mergers, see de
Lucia \etal 2006 and Kaviraj \etal 2005), $\alpha$/Fe increases with mass
and the age-mass relation (see \S3.7).  While it is plausible to have all
the galaxy components with the same epoch of initial star formation, our
models of chemical enrichment require a varying mass to produce varying
metallicities.  Thus, hierarchical formation is an inadequate explanation
for these oldest cluster ellipticals.

\subsection{Gap Population}

Common to seven of our eight clusters is a second, younger population (the
exception is A119).  This second population is distinguished by a clear
gap, varying in width of between one to two Gyrs, separating the younger
population (which we will refer hereafter as the gap population) and the
oldest galaxies.  While this gap population is not as numerous as the
oldest galaxies, it is significant.  When we divide all galaxies with
metallicities greater than [Fe/H]=-1.7 (the limit of our models) into three
bins of greater than 10 Gyrs (591 galaxies, 53\%), between 7 and 10 Gyrs
(431 galaxies, 39\%) and less than 7 Gyrs (88 galaxies, 8\%), we see that
the gap population constitutes about 1/3 of all cluster galaxies.

The oldest galaxies, having ages near the age of the Universe, must be
objects with simple star formation histories, i.e. a short initial burst
followed by a quiescent history.  However the gap population, with a mean
age between 8 and 9 Gyrs, can reach this stage through various paths.
First, it must be remembered that the calculated age is a luminosity
weighted mean value.  For example, a mean age of 8 Gyrs can be obtained by
a late epoch of star formation (5 Gyrs later than the oldest galaxies) or
by an extended duration of star formation, i.e. going from 12 Gyrs ago up
to 4 Gyrs from the current epoch to average as 8 Gyrs).  A comparison of
just those two exact scenarios is shown in Figure 6, where we linearly
combine SED models from 4 to 12 Gyrs (assuming equal mass for each age,
green line) and compare them to a single age 8 Gyr model.  As can be seen
in Figure 6, it is impossible, in practice, to distinguish these two paths
of star formation history simply by colors.

A secondary clock to determine the duration of initial star formation in a
galaxy can be estimated by the ratio of $\alpha$ elements to Fe (Greggio \&
Renzini 1983, Thomas, Greggio \& Bender 1998), since delayed enrichment by
Type Ia supernova raises the abundance of Fe relative to $\alpha$ elements
(such as Mg and Ca).  An increasing $\alpha$/Fe ratio with galaxy mass is
found by Thomas, Maraston \& Bender (2002), which is interpreted as
increasing initial star formation duration with decreasing galaxy mass
ranging from less than 0.5 Gyrs duration for the highest mass galaxies to 6
Gyrs in duration for galaxies around $10^9 M_{\sun}$.  These duration
values would produce mean galaxy ages consistent with the gap population
ages, although we note that while there is a slight decrease in age with
lower mass for the gap population (Figure 4), this trend is not as strong
as would be expected from the change in $\alpha$/Fe with mass (Gallazzi
\etal 2006).

\begin{figure}
\centering
\includegraphics[scale=0.95]{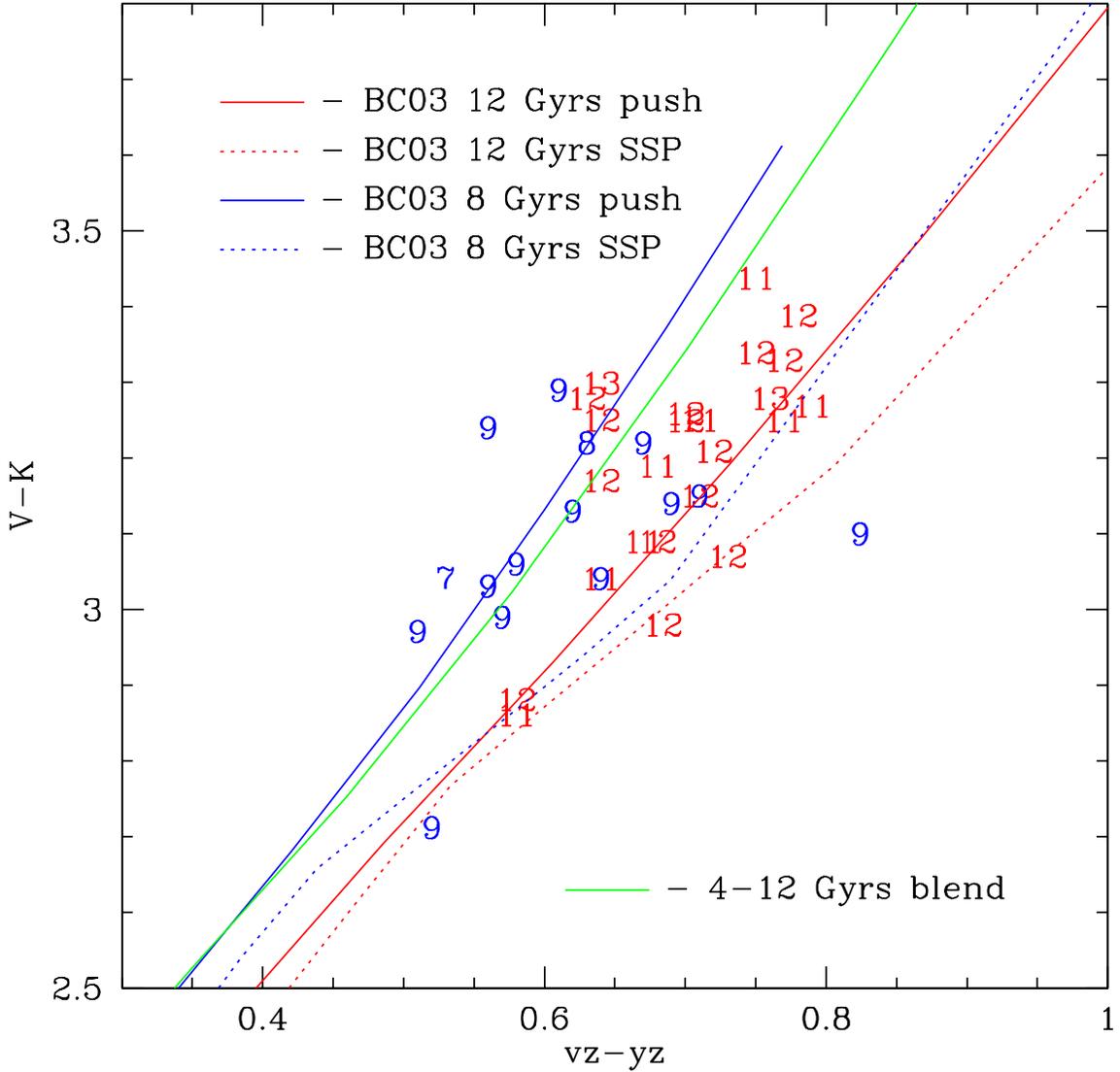}
\caption{Near-IR colors versus our narrow band color, $vz-yz$, for sixteen
Coma ellipticals.  Galaxies with ages less than 10 Gyrs are labeled by age
in blue, galaxies older than 10 Gyrs are labeled in red.  Two sets of 8 and
12 Gyrs tracks are shown from Bruzual \& Charlot (2003) SED models.  The
dotted lines are for SSP models, the solid lines are for our
multi-metallicity 'push' models (Schombert \& Rakos 2008) which blend in a
range of stellar metallicity populations using a simple infall model of
chemical evolution.  The 'push' models are an excellent match to the data
and the separation of age, as given by our continuum color technique, is
clear.  A blend model is shown in green, a sum of 16 models with ages
between 4 and 12 Gyrs in 0.5 steps to demonstrate that it is impossible to
distinguish between a galaxy with a singular age of 8 Gyrs versus the sum
of ages from 4 to 12 Gyrs.
}
\end{figure}

In addition to varying durations of star formation, an alternative
explanation to the younger age for the gap population is to consider the
effect of a more recent star formation event averaged with an old (12 Gyrs)
stellar population.  Recent burst models (referred to as frosting models,
Trager \etal 2000) are parameterized by the fraction of the mass of the
galaxy involved in the burst and when the burst occurred.  Obviously, the
effect of a star formation burst on the integrated color of a galaxy, will
be less if the burst involves a small fraction of the galaxy mass and/or
occurs farther in the past (the population having time to redden and blend
in with the older stars).  In order to constrain the magnitude of a
frosting star formation event, we can ignore any burst that would still
leave a signature of stars hotter than A type at the present day.  For
these stars, in sufficient numbers, would dominate the near-blue colors in
early-type galaxies, which have been known for decades to have the
continuum shape represented by K giants (O'Connell 1980).  However, a
population that is over one Gyr old has already taken on the colors close
to a metal-rich, old population and is only identifiable when mixed with an
older population using hyper accurate measurements, at least in the optical
region of the spectrum (Rose \etal 1994).

\begin{figure}
\centering
\includegraphics[scale=0.95]{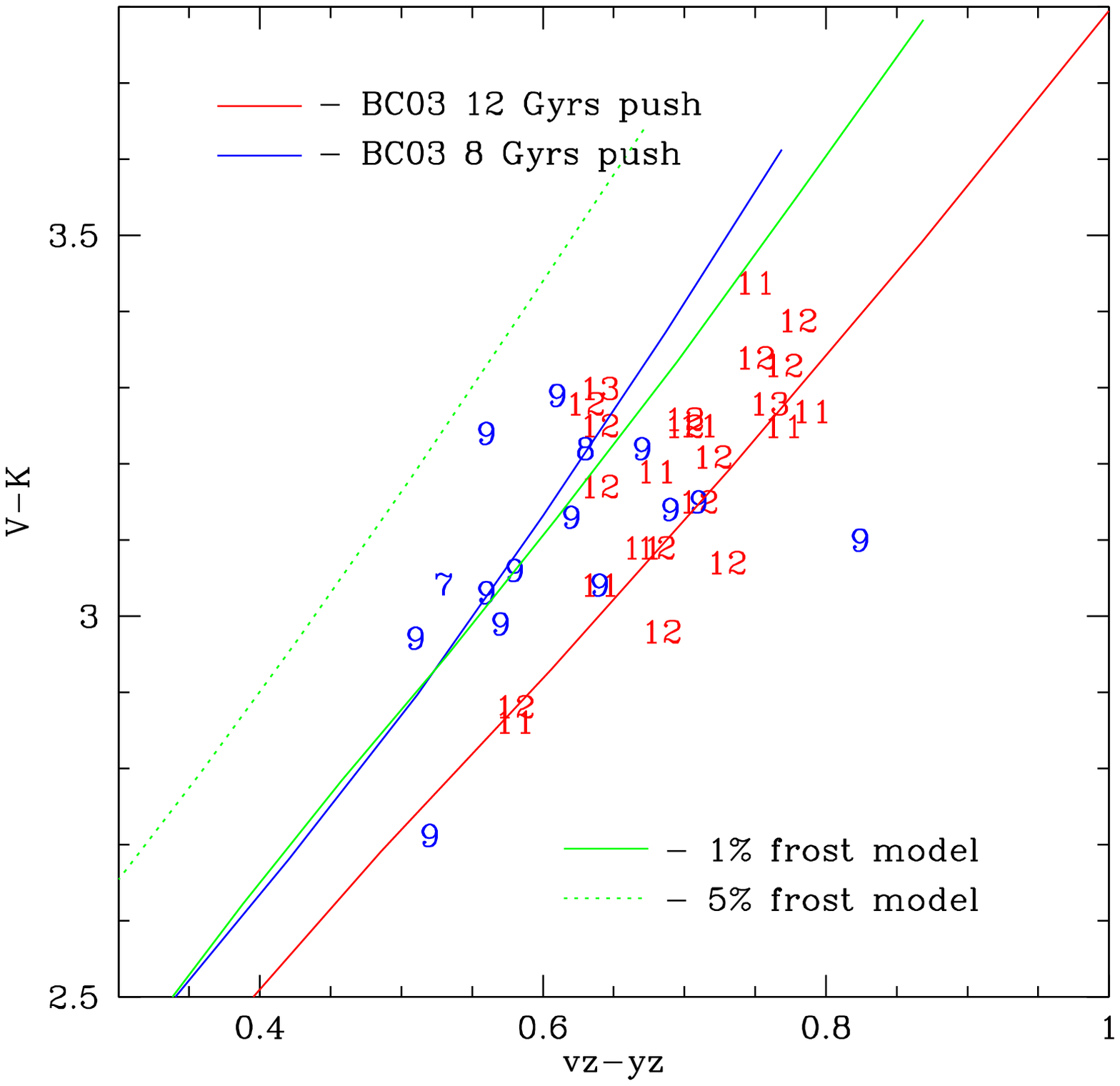}
\caption{Same as Figure 6, near-IR versus narrow band colors.  The same
infall models from Figure 6 are shown.  In addition, two 'frosting' models,
where a fraction of the galaxy mass is a population of 1 Gyr mixed with a 12
Gyrs population.  A 1\% frosting burst is nearly identical to a standard 8 Gyrs
population.
}
\end{figure}

To test the effect of an old, weak burst on the measured mean age of our
galaxies, we have mixed a 1 Gyr population of varying fractions to an old
12 Gyr population using the BC03 models.  Returning to the optical/near-IR
diagram we used in Figure 6, we have plotted a 1\% burst (involving 1\% the
mass of a galaxy) and a 5\% burst for a range of mean metallicities (again
using a metallicity distribution calculation as given by Schombert \& Rakos
2008).  These tracks are shown in Figure 7, again along with our Coma data.
From these tracks, it is clear to see that even a small burst of 1\% of the
mass of the galaxy, after 1 Gyr, has produced a trend of color which
closely resembles a composite 8 Gyrs population.  Stronger bursts, of up to
5\%, can be ruled out by the $vz-yz$ and $bz-yz$ narrow band colors (as
seen by the dotted line in Figure 7).  We conclude it is impossible to
distinguish between a small (less than 1\%), recent bursts or a composite 8
Gyrs model for the gap population.

\subsection{Age-Density Relation}

A sharp separation in age for cluster early-type galaxies suggests either a
formation process which delays or extends star formation for a subset of
galaxies (i.e. a nurture scenario) versus one where some environmental
process has operated on a subset of galaxies after the formation epoch to
induce later star formation (i.e. a nature scenario).  Environmental
processes are typically tested by looking for correlations with galaxy
density, on the assumption that the operating process is tied to the
dynamical interactions either with the cluster tidal field (galaxy
harassment, Moore \etal 1996) or by galaxy interactions.

While this environmental effect may be recent, as suggested by the frosting
models, this may also be a remnant of a primordial event.  For example, one
of the key predictions for hierarchical galaxy formation models is a
difference between field and cluster galaxy ages (where cluster systems
form first and, therefore, have older stellar populations).  Our sample is
strictly a sample of rich cluster galaxies, however, we can test for any
correlation with local density or with cluster radius, a crude measure of
local density.  Just such a test is found in Figure 8, a plot of galaxy
density as a function of cluster radius in kpc.

\begin{figure}
\centering
\includegraphics[scale=0.95]{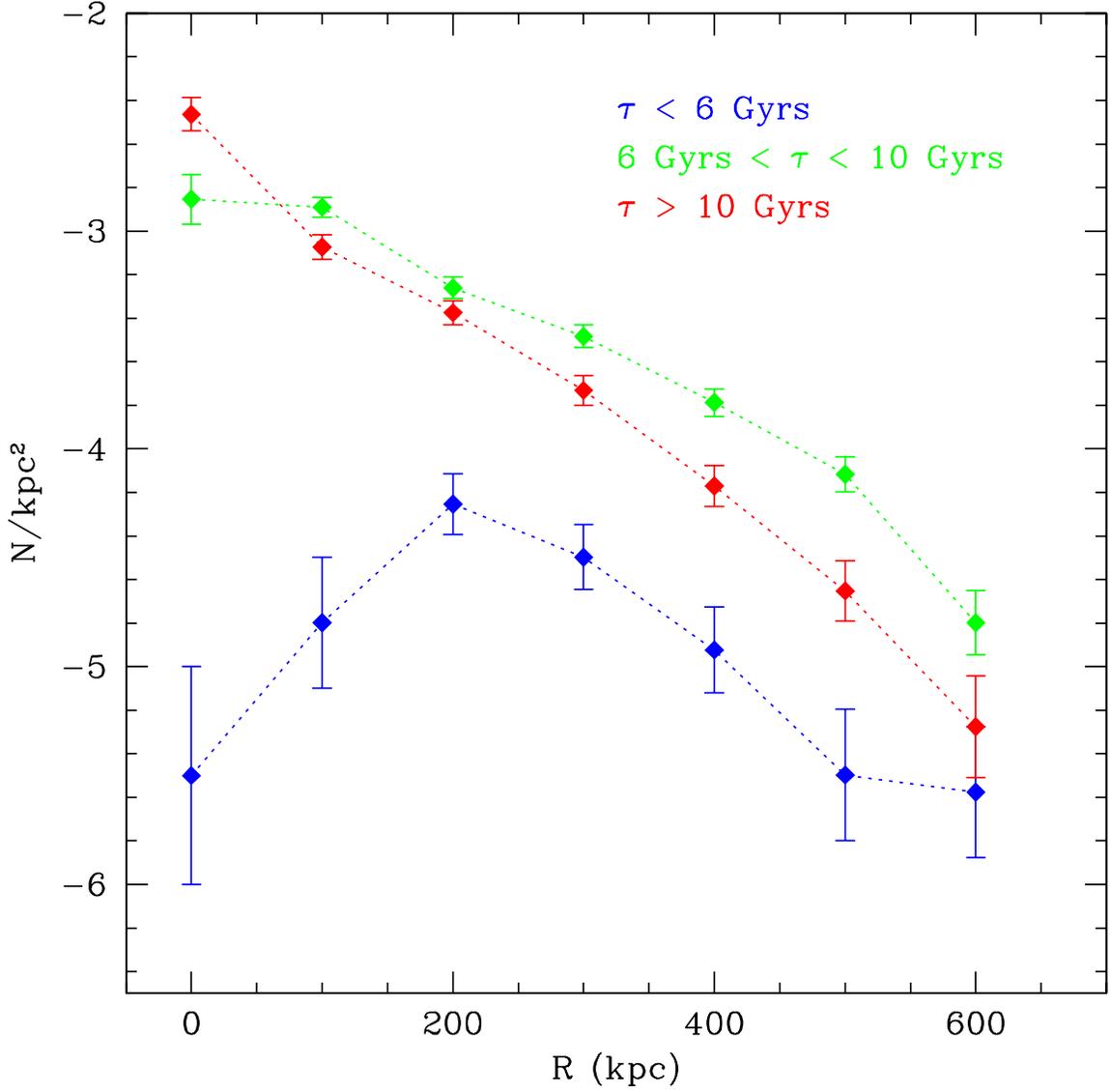}
\caption{Galaxy density as a function of cluster radius for galaxies older
than 10 Gyrs (red), between 6 and 10 Gyrs (green) and younger than 7 Gyrs
(blue).  Young galaxies clearly avoid the cluster center, however, even
intermediate aged galaxies are less clustered than the oldest galaxies.
Older galaxies are the most concentrated to the cluster center (higher
densities), all these trends suggest an environmental influence on mean
galaxy age.
}
\end{figure}

To produce this plot, we have binned all the clusters into one geography
(i.e. no correction for cluster size).  The cluster center is determined as
either as the position of the brightest cluster member, or the geometric
center of the cluster weighted by galaxy mass.  A119 was excluded in the
calculations as it does not contain a gap population and A1185 was excluded
since its irregular structure made the determination of the cluster center
problematic.  The resulting sample of 652 galaxies were binned in 100 kpc
radii and the number of galaxies per annulus area were summed for galaxies
with ages greater than 10 Grys (above the gap), between 7 and 10 Gyrs
(below the gap) and less than 7 Gyrs.

Immediately obvious from Figure 8 is that galaxies with younger mean age
avoid the denser regions of the clusters and the older galaxies are more
concentrated.  It is not surprising to find the very youngest galaxies
located on the cluster outskirts as these objects are structurally similar
to disk galaxies (see \S3.6) and have similar properties to the
Butcher-Oemler population.  However, even the intermediate aged galaxies
are less concentrated than the oldest cluster members.  This effect, that
younger galaxies avoid the cluster core, is slightly correlated with
cluster type, the more evolved cluster types display a stronger separation
by age than clusters that are dynamically young (as expressed by the BM and
RS types), leading to the obvious conclusion that dynamical processes
influence a galaxy's star formation history.  It is also notable that the
highest redshift cluster in our sample (A2218) displays this separation to
the greatest effect suggesting a process that has evolved recently in
lookback time.

\begin{figure}
\centering
\includegraphics[scale=0.95]{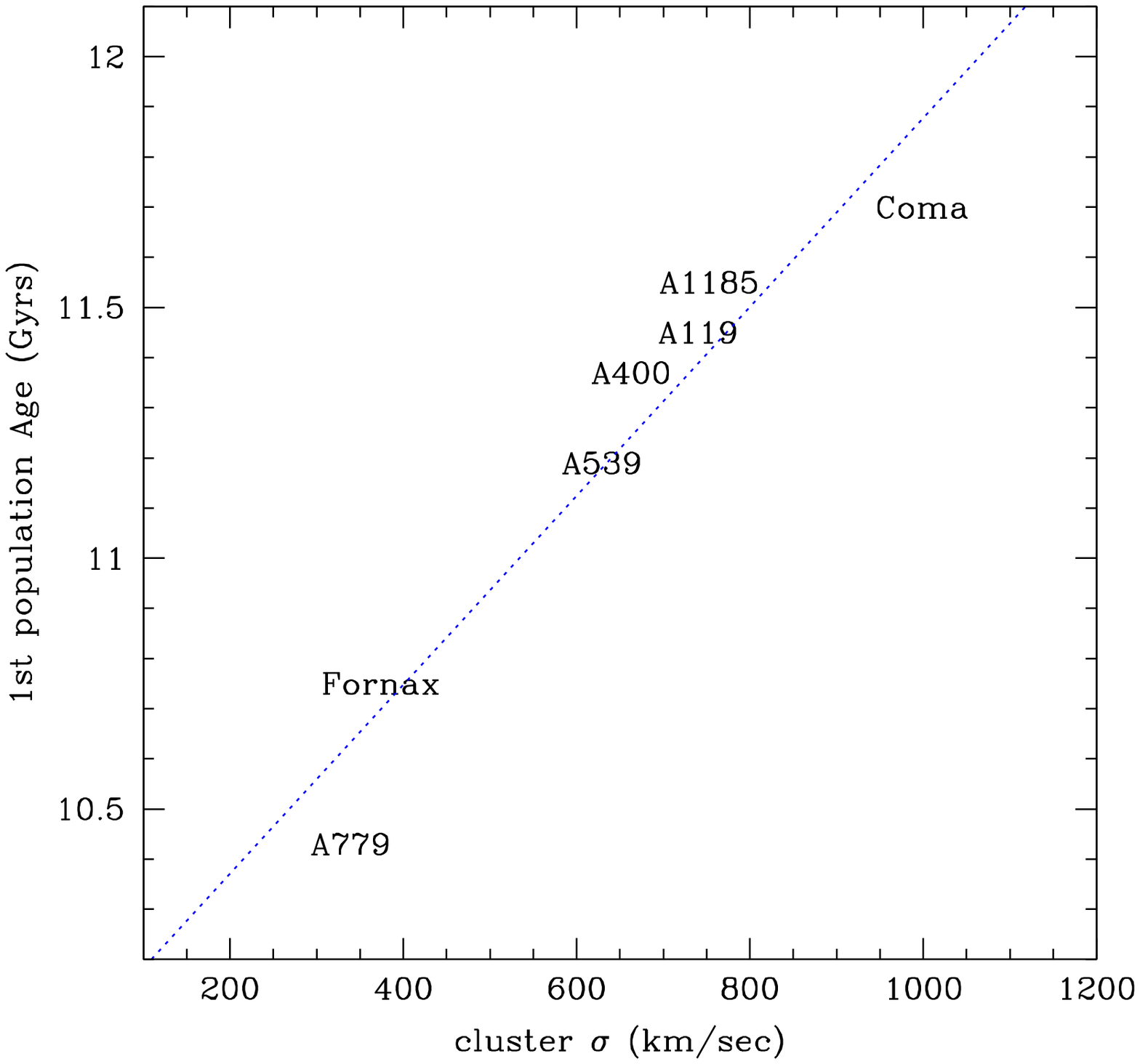}
\caption{Cluster velocity dispersion versus the mean age of the oldest
galaxies in each cluster.  The clusters with the highest velocity
dispersions (i.e. highest mass) have the oldest galaxies, implying earlier
cluster formation epochs.
}
\end{figure}

Environmental effects are not limited to the gap population.  Figure 9
displays the average age of the oldest galaxies (the 1st population) as a
function of cluster velocity dispersion.  Although the sample size is
small, there is a clear correlation between the mean age of the oldest
galaxies and dynamical state of the cluster as represented by velocity
dispersion.  The correlation here is such that high mass, high velocity
dispersion clusters have the oldest 1st populations, implying that these
clusters started galaxy formation (and the initial epoch of star formation)
before their lower mass counterparts by approximately 1 Gyr.

\subsection{Age-Morphology Relation}

A correlation between galaxy age and local density would imply a
correlation between galaxy type (morphology) and age by way of the cluster
density-morphology relationship (Ball \etal 2006).  To test this hypothesis
we extracted morphological types and structural parameters for the Coma
sample (see Odell, Rakos \& Schombert 2002).  Visual morphology was divided
into four types, $r^{1/4}$ objects (ellipticals), objects with distinct
bulge and disk components (S0's), objects with spiral or irregular
structure (Sp/Irr) and objects with lenticular or dwarf appearance (dE's).
Structural morphology was based on surface photometric fits to objects
brighter than $-$18.  These fits are divided into three types, bulge
objects ($r^{1/4}$), bulge+disk objects (S0's) and exponential disks
(lenticulars).

\begin{figure}
\centering
\includegraphics[scale=0.95]{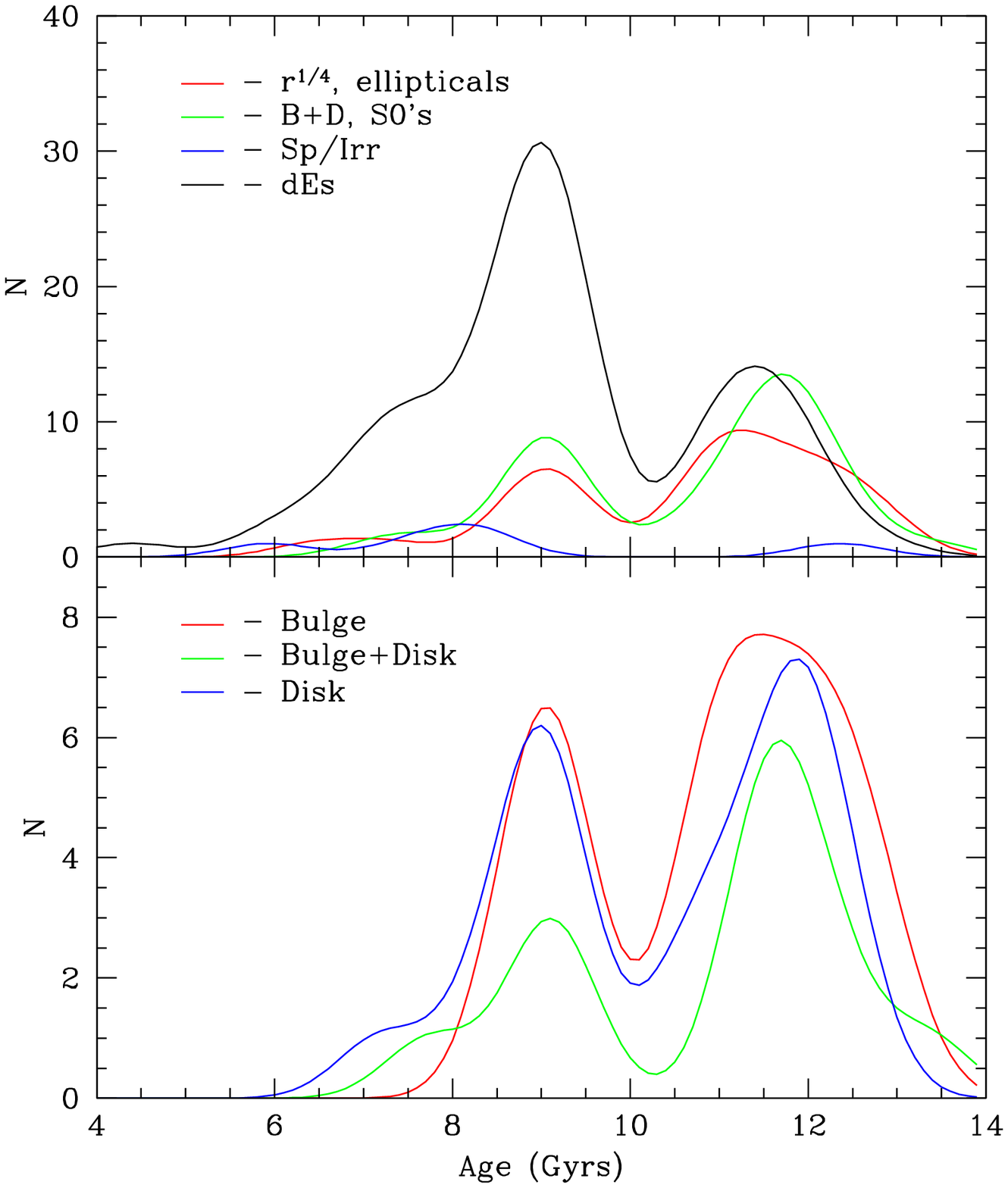}
\caption{Histograms of the ages of various morphological and structural
types in Coma.  Visual morphology was divided into four types, $r^{1/4}$
objects (ellipticals), objects with distinct bulge and disk components
(S0's), objects with spiral or irregular structure (Sp/Irr) and objects
with lenticular or dwarf appearance (dE's).  Structural morphology was
based on surface photometric fits to objects brighter than $-$18.  These
fits are divided into three types, bulge objects ($r^{1/4}$), bulge+disk
objects (S0's) and exponential disks.  Aside from a tendency for young
galaxies to be lower in luminosity (therefore, a lenticular or dwarf
elliptical appearance), various ages are found for all morphological and
structural types.   Surprisingly, S0's do not have a different age
distribution from ellipticals despite the expectation of quenching at
intermediate redshifts.
}
\end{figure}

Histograms of both visual and structural morphology are shown in Figure 10.
Surprisingly, there is no clear trend in morphology with age, such that the
percentage of galaxies classed E:S0:dE:S/Irr less than 10 Gyrs in age is
12\%:14\%:65\%:9\% versus greater than 10 Gyrs of 29\%:34\%:34\%:3\%.
Elliptical and S0 type galaxies have the same distribution in age, a
majority being old ($\tau$ = 12 Gyrs), but also form a significant fraction
in the gap population.  Star forming spirals and irregulars tend to be
young, but a majority of the young galaxies in Coma are in the lenticular
or dwarf category.  These morphological types are simply low luminosity
ellipticals with exponential shapes characteristic of ellipticals less than
$-$18 (Schombert 1987).  Their young age reflects the age-mass relation
discussed in \S3.1.

In terms of structure, the fraction of young to old galaxies are similar
for galaxies with bulges (ellipticals), bulge+disks (S0's) or pure disk
systems (lenticulars).  The fractions B:B+D:D with ages less than 10 Gyrs
is 37\%:21\%:42\% versus the fraction for galaxies with ages greater than
10 Gyrs of 44\%:24\%:32\%, i.e. slightly more r$^{1/4}$ objects over pure
disk systems for older galaxies.  Since the structural sample is composed
solely of the brightest galaxies in Coma, this histogram reflects the lack
of any morphological distinction in age for the top of the cluster
luminosity function.  Whatever processes determines a galaxy's age is
decoupled from its morphological evolution (outside of current star
formation).

The lack of an age difference between ellipticals and S0's (or bulge
objects versus disk lenticulars) is surprising as there is an expectation
that S0 disks are only recently quenched (i.e. stripped of their gas) and
should display a younger age (Bedregal \etal 2006, Burstein \etal 2005).
Although we note that the largest B/D S0's are the oldest such that even
the oldest S0's may have young disks buried in the light of their dominant
bulges.

\subsection{Age Correlations with Mass and Metallicity}

The existence of correlations between a galaxy's stellar population age and
mass or mean metallicity has been debated in the literature (Conselice
2007), but several recent large studies have converged on a consistent
picture (Thomas \etal 2005, Gallazzi \etal 2006, Jimenez \etal 2006,
Annibali \etal 2007).  While differing in the details, all these studies
find that galaxy age increased with stellar mass (luminosity) and
metallicity ([Fe/H]), although the deduced correlations contain a large
amount of uncertainty.  And, common to all these studies, the scatter in
the correlations exceed the observational errors, implying secondary
processes are significant.

For this study, noting again that we measure age and metallicity based on a
different technique than line indices, we find the same general
correlations between stellar mass (luminosity) and metallicity as previous
work (Figures 2, 3 and 4).  While the correlation between stellar mass and
metallicity ($<$Fe/H$>$) is evident, the scatter, like the line indices
studies, is clearly greater than the observational uncertainties and,
therefore, the depth of the gravitation well of a galaxy is not the sole
parameter that drives the chemical evolution of a galaxy (as predicted by
galactic wind infall models, Pipino \& Matteucci 2006).  Dividing the
sample by age (for example, considering the mass-metallicity relation
solely for galaxies with ages greater than 10 Gyrs), we find that the oldest
systems display a tighter mass-metallicity relation, however continue to
display scatter larger than observational errors.

In terms of age, there is a distinct separation by mass between the old
galaxies and the gap population.  For galaxies brighter than $M_{5500} =
-18$, the breakdown between the old galaxies and the gap population is 68\%
to 32\%.  For fainter galaxies, this division reduces to 49\% to 51\%.  But
while there is a tendency to find the youngest galaxies to be of the lowest
masses, there is no clear linear correlation.  This is due, primarily, to
the fact that the division into two populations is so distinct that each
population should be analyzed separately.

Considering the oldest galaxies first, this population has a mean age of
11.5 Gyrs.  To within the uncertainties, the slope of the old population
with respect to stellar mass is very shallow with a change of less than 1
Gyr over the full range in mass.  For the younger gap population, there are
larger numbers of young galaxies at low masses ($M < 10^9 M_{\sun}$).  This
agrees with the shape and scatter of the color-magnitude relation (Odell,
Rakos \& Schombert 2002), however, some of these galaxies probably
represent the low mass Butcher-Oemler objects (Rakos \etal 1996) with
traces of recent star formation (see Figure 13).  Likewise, the small
number of high mass galaxies ($M > 10^{10} M_{\sun}$) with ages less than 8
Gyrs could easily represent cluster galaxies which have undergone recent
mergers of disk galaxies with younger stellar populations or induced star
formation.

The age versus metallicity diagram (Figure 4) displays a wealth of buried
information.  The old population of galaxies shows a slight trend for an
increase in $<$Fe/H$>$ with age, certainly the oldest galaxies ($\tau$ $>$
12 Gyrs) all display greater than solar metallicities.  But the trend is
weak and galaxies with an age of 11 Gyrs have a range of [Fe/H] from $-$1.2
to $+$0.3.  The implies that the age of the stellar population does not
have a large influence on chemical evolution and that the final metallicity
of a galaxy is dominated by its mass.

The situation differs for the younger gap population.  Here we find a
fairly good correlation between age and metallicity such that the older gap
galaxies ($\tau = 9$ Gyrs) have the highest metallicities (near solar).
The younger gap galaxies ($\tau = 7$ Gyrs) have lower metallicities, near
[Fe/H]$=-1$.  We note that this trend of age and metallicity for the gap
population is opposite to the correlation of age and metallicity (i.e.
younger galaxies having higher metallicities) demonstrated by Trager \etal
(1998), Jorgensen (1999), Ferreras \etal (1999), Terlevich \& Forbes (2002)
and S\'{a}nchez-Bl\'{a}zquez \etal (2006).  However, an extensive study of
the correlated errors by S\'{a}nchez-Bl\'{a}zquez \etal finds no such
relationship in high density regions.  As our cluster sample is extracted
from the very highest density regions in the Universe, it is no surprise
that our oldest galaxies display no correlation between age and
metallicity.

Our gap population displays the opposite correlation as in found for field
galaxies by the S\'{a}nchez-Bl\'{a}zquez \etal work, which we interpret to
mean that the measured age from these galaxies is determined by post
formation processes unique to the cluster environment.  If the younger age
of these galaxies is due to longer durations of initial star formation,
then we would expect the opposite correlation as more star formation pushes
towards higher metallicities.  Instead, the effect we find would support a
frosting scenario such that younger age correlates with mass (smaller
galaxies being more susceptible to a burst event), but that chemical
evolution is primordial and independent of measured age.

\subsection{Age Gradients}

One of the most common criticisms of the Lick/IDS technique is that it
finds a high fraction of galaxies with line indice determined ages of less
than 7 Gyrs, yet very red optical to near-IR colors. This is inconsistent
with other observations of ellipticals from the Fundamental Plane and at
intermediate redshifts (Rakos \& Schombert 1995).  A re-analysis of the
Trager \etal dataset by Schiavon (2007) finds that the number of young
ellipticals ($\tau <$ 7 Gyrs) had been overestimated.  Using an improved
set of model tracks, Schiavon finds 3/4's of the Trager \etal sample to
have mean ages greater than 7 Gyrs and only 1/4 younger.  This study finds
only 7\% of cluster galaxies younger than 7 Gyrs, so even the re-calibrated
Lick/IDS method still finds a higher fraction of young systems.  However,
the Trager \etal sample covers a range of galaxy environments, whereas our
data is strictly a cluster sample.  There are many studies suggesting an
age difference between the field and cluster environments (Thomas \etal
2005, S\'{a}nchez-Bl\'{a}zquez \etal 2006, Annibali \etal 2007) in the
direction of younger ages in their field samples.  But even considering
only their high density samples, we still do not find the wide range of
ages suggested by studies using the Lick/IDS system.

Another comparison to the age and metallicity values extracted for the
Lick/IDS system, with respect to our continuum color technique, can be made
by examining the SDSS averaged spectra produced by Eisenstein \etal (2003).
These spectra consist of thousands of SDSS spectra selected by color and
morphology.  These averaged spectra are divided into four luminosity bins
and, for our comparison, we use the spectra taken over the full range of
environmental densities.  Eisenstein \etal publish their Lick/IDS values
(which can be converted into age and metallicity through the MgFe and
H$\beta$ indices using BC03 models) and their spectra were re-reduced by
Schiavon (2007) using a different technique to derive the same Lick/IDS
indices.  Figure 11 displays the age-metallicity-stellar mass plane for our
sample and the Eisenstein SDSS spectra.  Our data are averaged in galaxy
mass bins of 0.5, shown by the black symbols in Figure 11.  Each data point
is the average of the bin, the error bars display the dispersion within
each bin (i.e. not uncertainties).  Note that the separation into two
galaxy populations by age is blurred by this type of analysis, but does
reproduce many of the correlations seen in line indices study (e.g.
increasing age with galaxy mass).  The blue diamonds display the original
Eisenstein \etal Lick/IDS values converted into age and metallicity.  The
red diamonds are the re-analysis values by Schiavon.  The green diamonds
are the spectra convolved through our narrow band filters and converted
into age and metallicity by comparison to our galaxy sample (the $uz$
filter is estimated from extrapolation of the near-UV portion of the SDSS
spectra).

\begin{figure}
\centering
\includegraphics[scale=0.95]{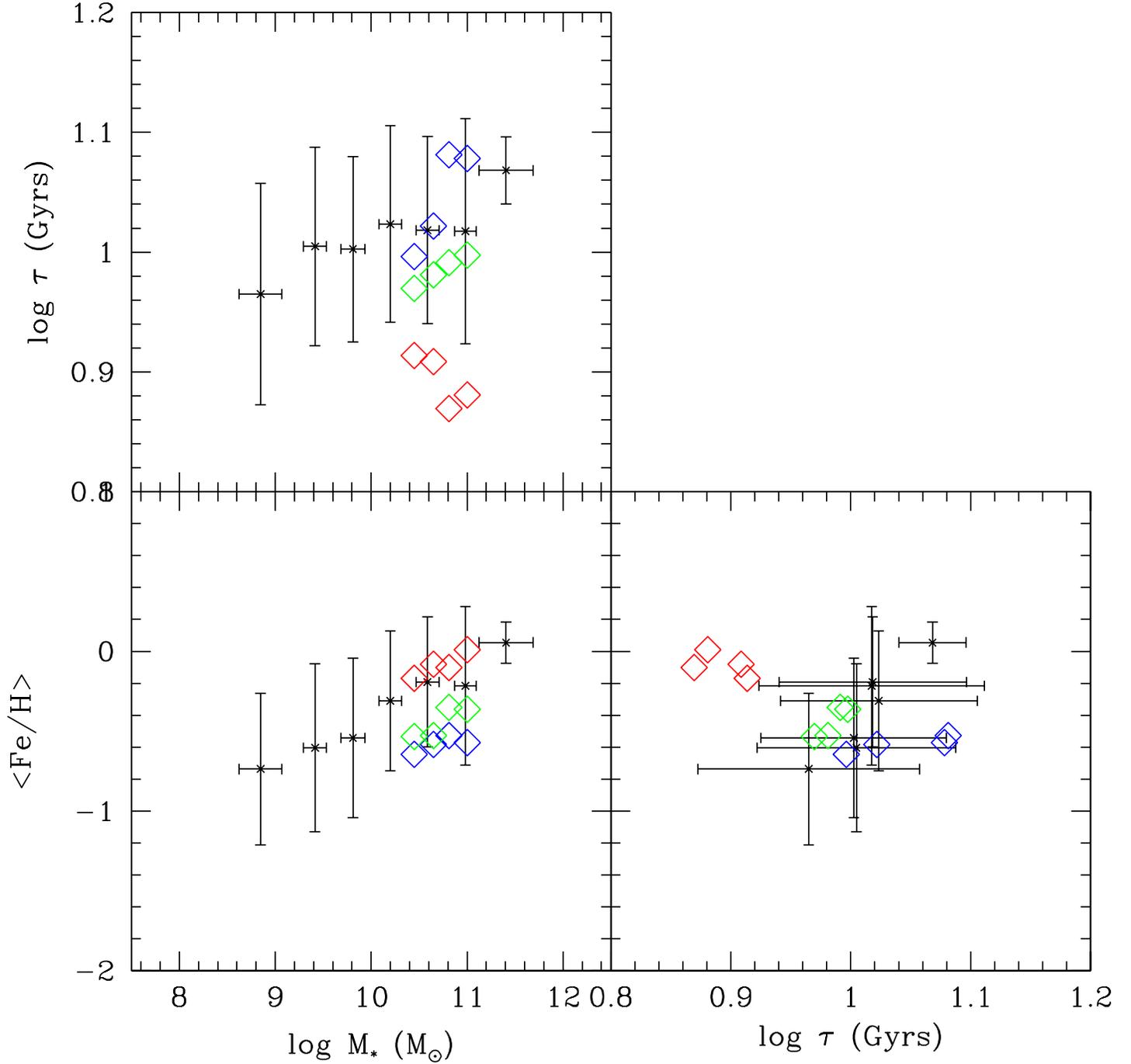}
\caption{Figures 2-4 redrawn with averaged values for our cluster sample.
Error bars display the dispersion in the averages, not uncertainties.  This
analysis blurs the dual nature of galaxy populations, but reproduces the
correlations seen in other studies.  Blue diamonds display the model
extracted values from the Eisenstein \etal SDSS averaged spectra.  The red
diamonds are Schiavon's (2007) re-analysis of those spectra using a
different Lick/IDS technique.  The green diamonds are the ages and
metallicities deduced from our continuum color analysis to the same spectra
(see text for discussion).
}
\end{figure}

Figure 11 displays some of the more common problems with the interpretation
of Lick/IDS values.  For example, the raw Eisenstein \etal values are in
fair agreement with our age values as a function of mass (blue diamonds),
but metal poorer than our average [Fe/H] values.  The Schiavon re-analysis
recovers metallicity, but now the calculated ages are several Gyrs younger
than our mean ages.  This is the same behavior discovered by Monte Carlo
simulations of correlated errors between age and metallicity by Thomas
\etal (2005).

When we reduce the Eisenstein spectra through our continuum color scheme we
recover the mean ages and metallicities of our cluster sample (although
slightly younger and more metal poor due to the mixture of field and
cluster spectra in the sample).  While the claim is made that the Lick/IDS
H$\beta$ index is age sensitive independent of abundance changes, the
comparison in Figure 11 appears to indicate otherwise.  The source of the
difference between the Eisenstein \etal Lick/IDS values and the Schiavon
analysis is the amount of in-fill corrections made to the spectra.
Schiavon himself notes that, without these corrections, the H$\beta$ ages
would be approximately 14 Gyrs, closer to our values.  Secondary age
estimates from the Lick/IDS system (such as the H$\delta$ index) produce
even younger mean ages, however, these indices have serious technical
difficulties in their application (Prochaska \etal 2007).

In section \S3.4 we showed that it is impossible to distinguish, using our
colors, between a galaxy with a mean age of 8 Gyrs and one with an old
population (13 Gyrs) mixed with a 1\% 1 Gyr burst.  However, the former
scenario would produce ages from the Lick/IDS system in concordance with
our values and the frosting scenario would be an obvious candidate to
explain the discrepancies between our age estimates and ones produced by
the Lick/IDS technique (Tantalo \& Chiosi 2004, James \etal 2005).
Schiavon demonstrates that Eisenstein \etal values could be reproduced by
assuming a 0.5 to 1\% burst with a 1 Gyr population, and exact match to the
effects of a mild burst on our continuum colors (see Figure 7).  To see how
a frosting scenario would resolve age discrepancy, consider the primary
differences between the results derived from spectroscopy (i.e.  the
Lick/IDS system) and our continuum color technique: 1) spectroscopic values
for [Fe/H] tend to higher than our values and 2) spectroscopic ages tend to
be younger than our calculated ages (see red versus green diamonds for
[Fe/H] in Figure 11 and the age histograms in Figure 5). In our most recent
study of the metallicity values for ellipticals (Schombert \& Rakos 2008),
we performed a detailed study of the differences in mean [Fe/H] as
determined by the Lick/IDS system (i.e. a direct measure of Fe lines)
versus our technique (which determines [Fe/H] by model interpretation of
the colors as driven by the position of the RGB).  The correspondence
between the two metallicities derived by spectroscopy and colors was good,
however, the Lick/IDS values were consistently higher than our estimates by
approximately 0.2 dex.

Examination into how the Lick/IDS measurements are made revealed the source
of the difference in our two metallicities.  Being a spectroscopic
technique, the Lick/IDS values are extracted by slit observations of
typically round galaxies.  This means the light observed by spectroscopic
methods is light extracted from a slice through the galaxy's core.  For
galaxies in the redshift range of the Trager \etal study (between 1,000 and
5,000 km/sec), this means that the light through the slit represents only
1/10 to 1/3 the light of a galaxy as compared to our total luminosities or
aperture colors (for standard slit widths of between 2 to 5 arcsecs).  In
addition, the light obtained by spectroscopy is highly weighted by core
luminosity.  In the typical galaxy observed at the distance of Coma, 70\%
of the luminosity measured by slit spectroscopy is due to stars within 1
kpc of the center (versus only 20\% for imaging).  For fiber studies, this
effect is magnified as the data will only encompass the core of a galaxy
with no halo component.  When combining this bias with typical color
gradients, Schombert \& Rakos (2008) estimated that the resulting Lick/IDS
metallicity values would be approximately 0.2 dex higher than values
deduced from all the galaxy light (using a simple model of internal
metallicity distribution), exactly in agreement with the color deduced
values.

Whether the difference in age measurements made by the Lick/IDS system,
versus our continuum color technique, is due primarily to an aperture
effect will depend on the degree in which age gradients are found in
ellipticals.  Color gradient work finds that the measured gradients in
ellipticals are fully explained by changes in metallicity from galaxy core
to halo (Thomas \etal 2005).  This is also supported by a study of color
gradients as a function of redshift (Tamura \etal 2000) which found no
evidence for changes in color gradients due to lookback time out to a
redshift of 1.  On the other hand, several line indice studies have
detected significant age gradients in the H$\beta$ index (Tantalo \etal
1998, S\'{a}nchez-Bl\'{a}zquez \etal 2006b).  Both these studies found
younger stellar populations in their sample galaxy's cores, as well as
finding the cores to be more more metal rich and enhanced in $\alpha$
elements (which would imply star formation from a simple burst rather than
a prolonged event).

\begin{figure}
\centering
\includegraphics[scale=0.85]{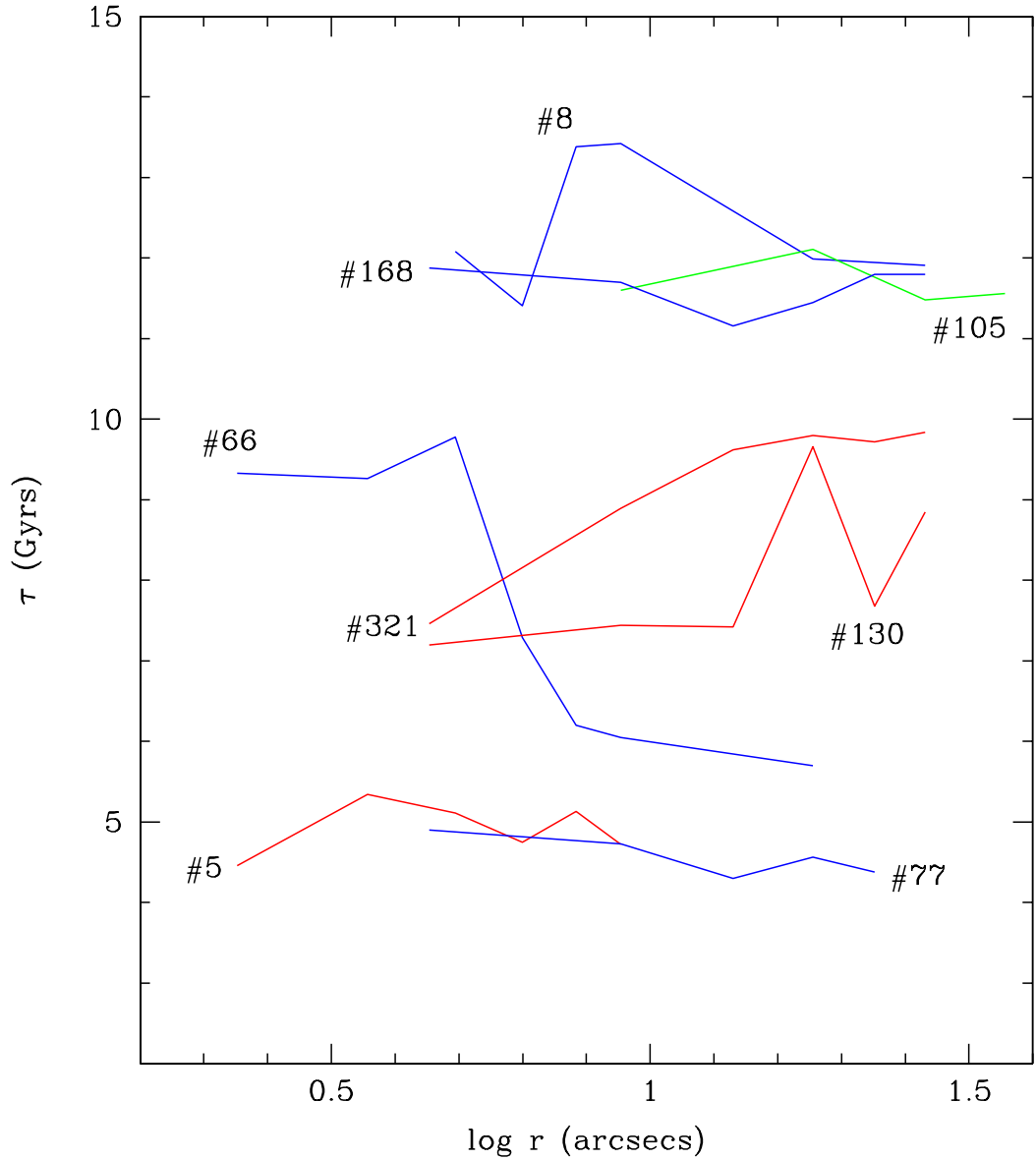}
\caption{Age gradients for eight galaxies in A1185.  Shown are age
estimates for various integrated apertures listed by log radius.
Decreasing age gradients are shown in blue, increasing age gradients are
shown in red, neutral in green.  While galaxies with younger ages are found
in A1185, their gradients are insufficient to explain the young galaxy ages
measured by line indices studies.
}
\end{figure}

Thus, the discrepancy between ages determined by our continuum colors,
which measure the integrated light from a galaxy, can be resolved if one
considers a burst model.  As bursts of star formation are concentrated in
galaxy cores (Thomas 1999), then this will produce mean ages from
spectroscopy skewed towards the burst mean age.  In addition, the burst
population, being formed later in a galaxy's life, will be enriched in
metallicity (agreeing with metallicity gradients).

To consider age gradients in our own continuum color sample, we examined
the change in age with radius for a sub-sample of galaxies in A1185.  These
age values are displayed in Figure 12, where the aperture radius is used
and the calculated age values are deduced from the integrated colors of
each aperture (versus a differential color).  The observed gradient are
small, typically only a few Gyrs.  There are equal number of positive and
negative gradients in Figure 12 and certainly nothing to indicate that
spectroscopic studies are being influenced by an extremely young core
population.  However, as discussed in \S3.4, our intermediate ages may 
be the combination of a young (less than 1 Gyr population) and an old, 12
Gyr population.  Spectroscopic signatures of a burst population (i.e. the
Balmer lines) may dominate the our continuum colors (Serra \& Trager 2007).
If true, then this would mean that the Lick/IDS technique is a powerful
method of detecting small, medium aged bursts, but 1) produces an age
measurement for the entire galaxy stellar population that is biased towards
young values, 2) overestimates the metallicity of the galaxy (instead it
measures the core [Fe/H] which should be a maximal value for the galaxy)
and 3) deduces trends in $\alpha$/Fe which reflect the chemical evolution
of the burst population, not the global values for the galaxy as a whole.
All consistent with the differences we observe between color and line
indice estimates of age and metallicity.

Lastly, the age differences between colors and line indices may also be due
to other hot star contributions to a galaxy's color and spectrum.  For
example, metal-poor horizontal branch (HB) stars display spectral features
similar to new star formation signatures, particularly to the $H\beta$
index and near-blue colors.  However, in our examination of the internal
metallicity distribution for ellipticals (Schombert \& Rakos 2008), we
estimated the contribution from an old, metal-poor component with a
multi-metallicity model and measured its effect on our colors.  We found
that, while metallicity estimates must take an old population into account,
the metallicity distribution derived is similar to that proposed by infall
models of chemical evolution and that the calculated mean age is
insensitive to the metal-poor components (Schombert \& Rakos 2008).  This
is a recent change to our calculations due to the fact that the newest SED
models (Schulz \etal 2002) use isochrone tracks and properly fit the
advanced stages of stellar evolution (i.e. HB populations).  This result is
also in agreement with a similar analysis by Trager \etal (2005) where they
find that the high H$\beta$ values for Coma galaxies has a small component
from old, metal-poor HB stars, but that a younger burst population must be
mixed with an older, primordial population to account for the variation in
metallicity and age indices.

\section{SUMMARY}

We have presented a compendium of age and metallicity values determined
using a narrow band color system.  This method differs from age and
metallicity determination by spectroscopic methods (e.g. the Lick/IDS
system) by using a principle component analysis technique tied to SED
models and calibrated by galactic globular age and [Fe/H] values.  In
addition, we have confirmed our age values by comparison to near-IR colors
and varying SED models.  Degeneracy constrains the range of age and
metallicity of our method to objects greater than 3 Gyrs and more
metal-rich than [Fe/H]=$-$1.8, acceptable limits for the study of red
cluster galaxies.  Objects with current or very recent star formation can
be identified by their colors, but their age and metallicity are not
determined by our technique.

Our results can be summarized as the following:

\begin{itemize}

\item{} We find that cluster early-type galaxies divide into two distinct
populations.  An old population with mean ages similar to the age of the
Universe (i.e. primordial formation times).  A second younger population,
separated by a gap in age of about 2 Gyrs from the oldest galaxies.  The
age of the gap population is confirmed by their near-IR colors.  This
distribution differs compared to age distributions from SDSS studies (e.g.
Gallazzi \etal 2006), in that we find very few cluster galaxies with ages
less than 7 Gyrs and the gap is distinct in all our clusters.  In addition,
the lack of a smooth transition in age between the galaxy populations
suggests that a secondary process influences the measured age of galaxies
independent of the formation process.  The large number of old galaxies
will be a challenge to hierarchical models as the seed systems under those
formation scenarios will require identical ages and metallicities.

\item{} We cannot distinguish, using SED models, between star formation
history scenarios where the younger gap population is composed of a young
age due to 1) later formation redshift, 2) longer initial star formation
duration or 3) a later burst of star formation (a 'frosting event', Trager
\etal 2000).  However, if the gap population is responsible for the high
fraction of very young ages ($\tau <$ 7 Gyrs) found by many Lick/IDS
studies, then this would argue for a frosting model as the spectroscopic
studies are more sensitive to a younger population (Serra \& Trager 2007),
which would also explain the changes in $\alpha$/Fe with galaxy mass, a
secondary clock to measure star formation duration.

\item{} While the oldest galaxies display very little correlation between
age and metallicity, the gap population has a weak positive correlation
between mean age and $<$Fe/H$>$.  Since the youngest galaxies have the
lowest metallicities, then this implies that late star formation does not
contribute to the global metallicity of a galaxy.

\item{} A clear environmental imprint is found on the younger gap
population as its cluster distribution is less concentrated than the older
galaxies.  While a later epoch of formation would explain this distribution
(core galaxies forming a few Gyrs before the infalling population), the
clustering properties of the younger galaxies are similar to the higher
redshift Butcher-Oemler galaxies.   We interpret this correlation to imply
a dynamical evolutionary effect that induces late bursts of star formation.

\item{} The age of the oldest galaxies in a cluster is correlated with the
cluster's velocity dispersion (i.e. mass) which implies that galaxy and
star formation in the more massive clusters begins earlier than in less massive
clusters.

\item{} Surprisingly, there is no correlation between mean stellar
population age and galaxy morphology or structure.  Despite a large number
of S0 galaxies in our sample, there is no evidence that they have younger
mean ages as one would expected if their disks had been quenched in the
last few Gyrs.

\item{} Age gradients were examined for a sub-sample of galaxies in A1185.
Positive and negative age gradients were found with magnitudes of less than
a few Gyrs from core to halo.

\end{itemize}

A clearer picture begins to emerge concerning the ages of galaxies when we
combine the results from narrow band colors, near-IR colors, line indices
and high redshift work.  The correlations (and lack thereof) found herein,
between age, metallicity and galaxy mass, support previous conclusions from
the color-magnitude relation and galaxy velocity dispersion-metallicity
relations.  We also clarify the fact that the scatter in these relations
exceeds the expected uncertainties from observational error.  This is
demonstrated in Figure 13, the color-magnitude for our $vz-yz$ color
divided between the old population and the younger gap galaxies.  The
scatter is reduced dramatically when only the oldest galaxies are
considered, even though the effect of age on $vz-yz$ is negligible over the
range of 8 to 12 Gyrs.  This implies the path that a particular
galaxies takes through star formation and chemical evolution can vary,
particular within varying environments and the oldest galaxies truly
represent a standard monolithic collapse origin with a simple history of
chemical evolution.

\begin{figure}
\centering
\includegraphics[scale=0.85]{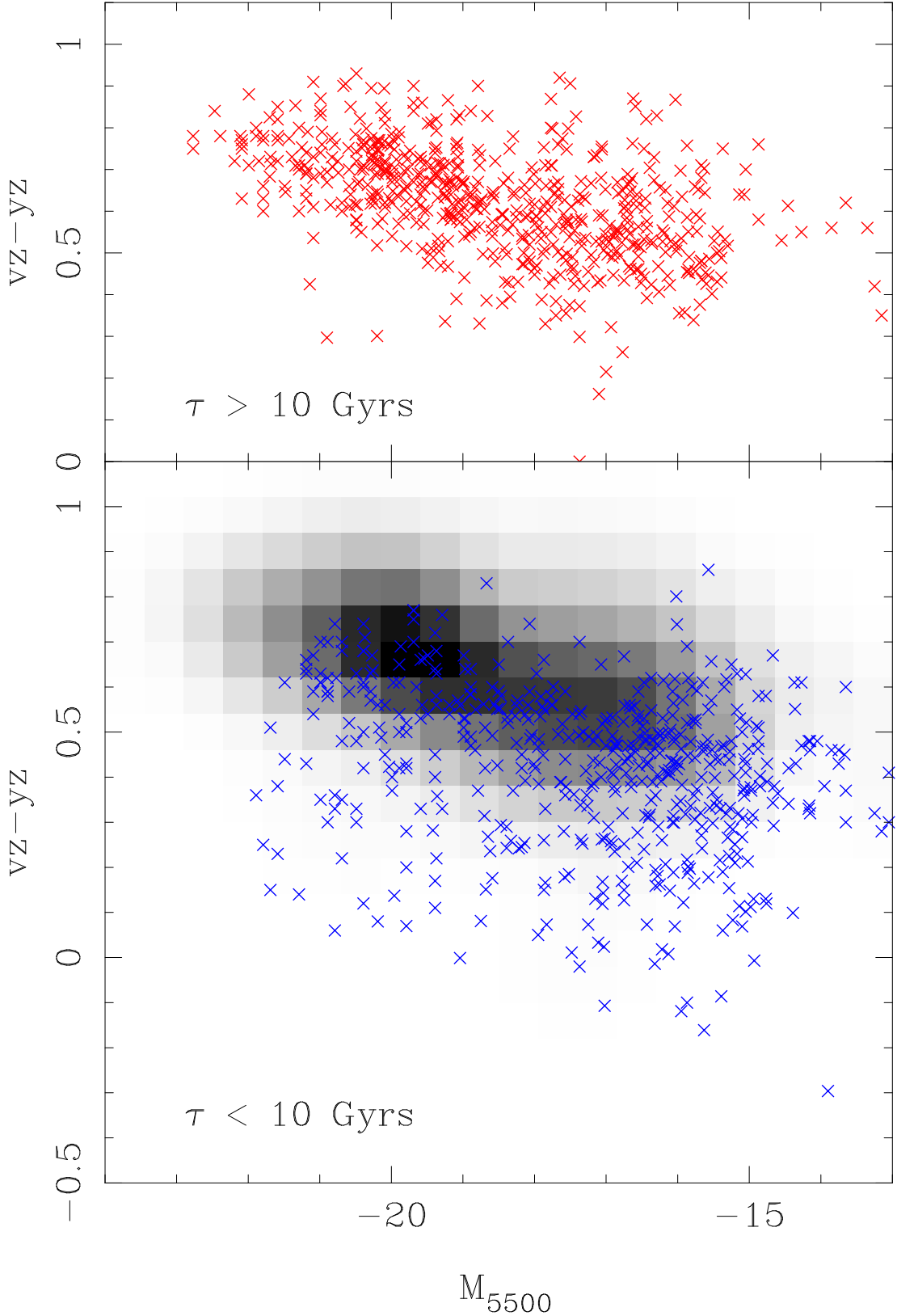}
\caption{The color-magnitude relation for our $vz-yz$ color (4100{\AA} -
5500{\AA}).  The top panel displays the CMR for all galaxies older than 10
Gyrs.  The bottom panel displays a greyscale of the top panel overlaying
all the data for galaxies younger than 10 Gyrs.  The scatter in the CMR is
notability reduced when just considering a uniform sample of galaxies
of singular age.
}
\end{figure}

A comparison between age and metallicity values determined by the Lick/IDS
system and our color scheme argues that the higher metallicities and lower
galaxy ages measured by line indices are real absolute values that reflect
different sensitives to age dependent features (see Serra \& Trager 2007).
Our technique finds a global value for age and metallicity that tends to be
less metal-rich and older in luminosity weighted age.  This would suggest
strong age gradients in early-type galaxies such that values from line
indice work are heavily weighted towards the core burst populations values;
however, none are detected in our sample and their varying effect on age
measurement techniques may disguise their presence.  

In conclusion, the study of age in galaxies suffers from the dual nature of
cluster galaxies in that they are are both young and old.  A majority of
cluster galaxies have primordial stellar populations, dating back to the
era of galaxy formation plus for all cluster galaxies, a majority of their
internal stellar populations are primordial.  However, a significant
fraction of cluster galaxies have a younger component that represents
between 1\% and 5\% of the galaxy mass.  The study of this burst population
illuminates recent environmental processes in clusters and provides the
connection to observed star formation at higher redshifts.

\acknowledgements

Financial support from Austrian Fonds zur Foerderung der Wissenschaftlichen
Forschung and NSF grant AST-0307508 is gratefully acknowledged.  We also
wish to thank the faculty and staff of the University of Arizona's Steward
Observatory for the time allocated to us using 90Prime on the Bok
Telescope.

\end{document}